%% file: aapps_HIreview.tex
\RequirePackage{fix-cm}
\documentclass[twocolumn]{svjour3}          

\usepackage{graphicx}   
\usepackage{xspace}     
\usepackage{url}
\usepackage[mathlines]{lineno}
\usepackage{bm}
\usepackage{amssymb, amsmath}
\usepackage{xcolor}

\usepackage[sort&compress,square,numbers]{natbib}
\usepackage{color}
\definecolor{orange}{cmyk}{0.,0.353,1.,0.}    

\begin{document}

\title{Signatures of QGP at RHIC and the LHC}

\author{T. \textsc{Niida}$^{1\dagger}$ \and Y. \textsc{Miake}$^{1,2\ddag}$}


\institute{
\at $^1$Department of Physics, Faculty of Pure and Applied Sciences, University of Tsukuba, Tsukuba, Ibaraki 305-8571 Japan. \and
$^2$Advanced Science Research Center, Japan Atomic Energy Agency, Tokai, Naka, Ibaraki 319-1195 Japan. \vspace{4pt}
\at
$^\dagger$niida@bnl.gov \\ $^\ddag$miake.yasuo.ge@u.tsukuba.ac.jp}

\date{Received: February 22, 2021 / Accepted: April 22, 2021}

\maketitle

\begin{abstract}
\input{0a-Abstract}
\end{abstract}


\section{Introduction\label{sec:intro}}
\input{1-Introduction}

\section{Jet quenching - High opacity -\label{sec:eloss}}
\input{2-EnergyLoss}

\section{Non-viscous flow\label{sec:hydro}}
\input{3-HydrodynamicalBehaviour}

\section{Other signatures and questions\label{sec:puzzle}}
\input{4-OtherSignaturesIssues}

\section{Where is the phase transition?}
\label{sec:PT}
\input{5-PhaseDiagram}

\section{Summary\label{sec:future}}
\input{6-Summary}

%
%
%
\begin{acknowledgements}
The authors thank Drs. S. A. Voloshin, Y. Tachibana, T. Hirano, T. Hatsuda, and S. Nagamiya for helpful discussions.  The authors also thank Drs. S. Esumi, T. Chujo, and H. Sako for daily discussions.
YM was supported in part by Japan Society for the Promotion of Science (JSPS) KAKENHI, Nos. 17H02876 25287048 and 20224014.
\end{acknowledgements}
%
%
\bibliography{0_references}
\end{document}

%% file: 0a-Abstract.tex
The progress over the 30 years since the first high-energy heavy-ion collisions at the BNL-AGS and CERN-SPS has been truly remarkable. Rigorous experimental and theoretical studies have revealed a new state of the matter in heavy-ion collisions, the quark-gluon plasma (QGP).
Many signatures supporting the formation of the QGP have been reported. Among them are jet quenching, the non-viscous flow, direct photons, and Debye screening effects.  In this article, selected signatures of the QGP observed at RHIC and the LHC are reviewed.

%% file: 1-Introduction.tex

For the study of extremely hot matter, experiments on ultra relativistic heavy-ion collisions have been carried out at the Relativistic Heavy Ion Collider (RHIC) at center-of-mass per nucleon-nucleon energies of $\sqrt{s_{\rm NN}} = 7.7-200 {\rm GeV}$ since 2000, and also at the Large Hadron Collider (LHC) at $\sqrt{s_{\rm NN}} =2.76-5.5{\rm TeV}$ since 2009.
Many experimental and theoretical studies have revealed a new state of matter, the quark-gluon plasma (QGP), in these collisions where quarks and gluons are no longer confined within hadrons. 

\begin{figure}[htb]
\begin{center}
\includegraphics[width=0.9\linewidth ]{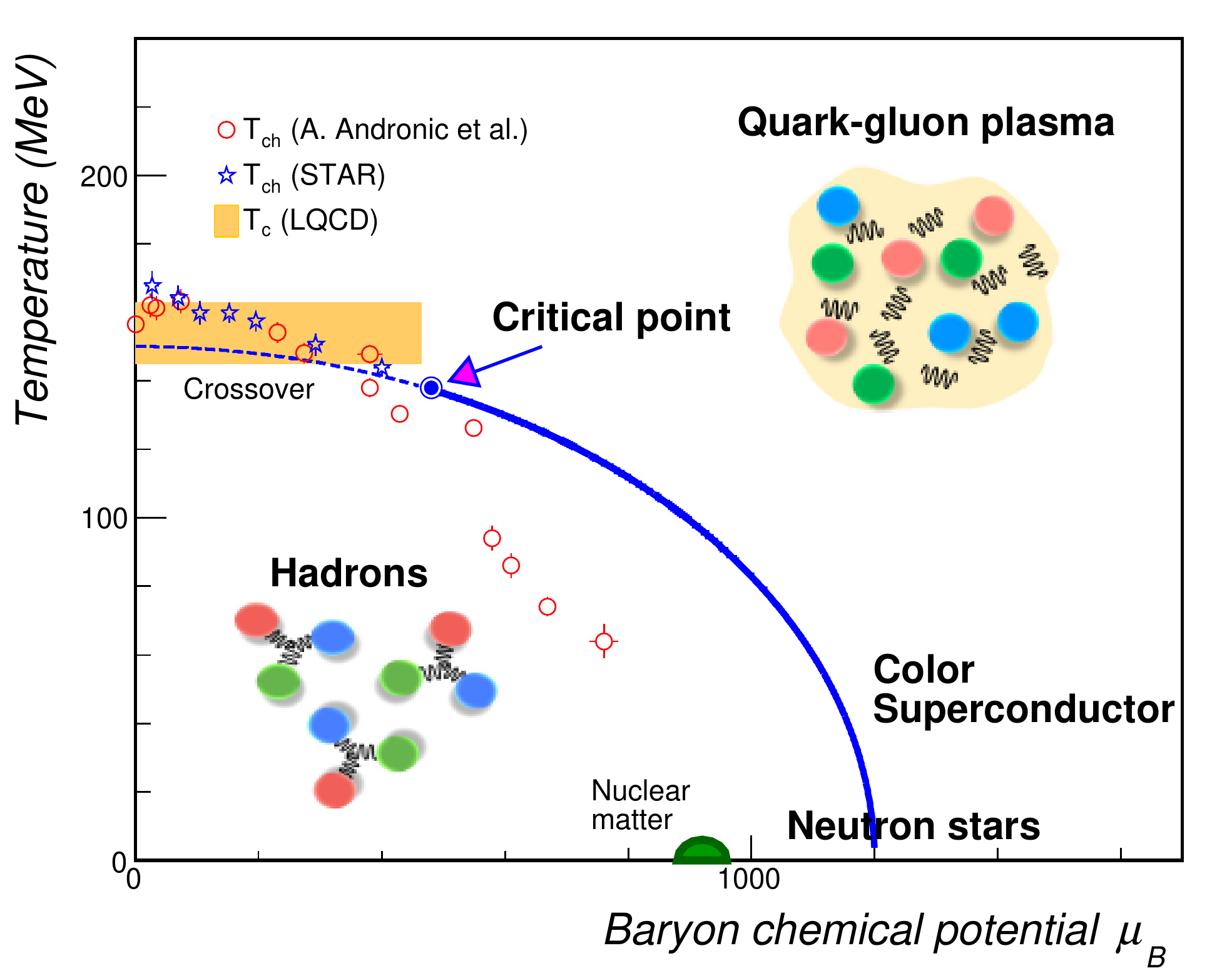}
\caption{A sketch of QCD phase diagram as functions of temperature and baryon chemical potential with conjectured phase boundaries indicated by solid and dashed lines and a possible critical point indicated by a solid circular point. Extracted $T_{\rm ch}$ and $\mu_B$ from a statistical model~\cite{Andronic:2017pug} are also plotted. See Sec.~\ref{sec:PT} for details.\label{fig:qcdPD}}
\end{center}
\end{figure}

Figure~\ref{fig:qcdPD} shows a schematic phase diagram of different phases for nuclear matter as functions of temperature and baryon chemical potential $\mu_{\rm B}$, with conjectured phase boundaries between the QGP and hadrons. 
Lattice QCD calculations predict a rapid but smooth crossover phase transition around the critical temperature $T_c\approx155$~MeV at small $\mu_{B}$~\cite{Aoki:2006we,Bazavov:2018mes}. Also, theoretical models suggest a first-order phase transition at high $\mu_B$ and the existence of the end point of the phase boundary, called the critical point. However the location of the critical point remains to be determined experimentally. 

Two distinct features were discovered at RHIC and then confirmed at the LHC: the high opacity of the matter and its non-viscous, fluidic nature.  The former, also known as jet quenching, is discussed in Section~\ref{sec:eloss} and the latter, the hydrodynamic behavior in Section~\ref{sec:hydro}. 
In Section~\ref{sec:puzzle}, other signatures supporting QGP formation together with related open questions are introduced.
Since the signatures of the QGP formation were observed both at RHIC and the LHC, the next question is where the phase transition exists.
In Section \ref{sec:PT}, the beam energy dependences of various observables are discussed in the context of the quantum chromodynamics (QCD) phase diagram.

There are many comprehensive review papers available~\cite{Schukraft:QM2017,Shuryak:2014zxa,PeterJohanna:2016,Science:2012,Muller:2006ee}. In this article, selected topics are reviewed, giving views on where we are and where we are going.


%% file: 2-EnergyLoss.tex
External probes are often utilized to see inside  matter and study its properties, e.g. the internal structure of hadrons by deep inelastic scattering. In heavy-ion collisions, the system life time is extremely short ($\sim10$~fm/$c$) and therefore it is almost impossible to use a literally external probe. Instead such a ``tomography" can be performed by  energetic and/or highly penetrating particles produced in initial parton-parton scatterings. The scattered partons have high transverse momentum and traverse the QGP losing their energy. Since the partons cannot exist on their own, they fragment into a spray of hadrons called a jet.

Energy loss of partons in the medium can be understood by collisional and radiative processes similar to energy loss in QED. Main difference between QED and QCD is that gluons interact with themselves unlike photons in QED. Collisional energy loss is due to elastic scatterings between the initial primary parton and a parton from the medium and is expected to linearly scale with the path length of the initial parton when traveling the uniform medium, while radiative energy loss takes place due to gluon radiation.

\begin{figure}[t]
\begin{center}
\includegraphics[width=0.44\linewidth,bb=0 140 1100 973,clip]{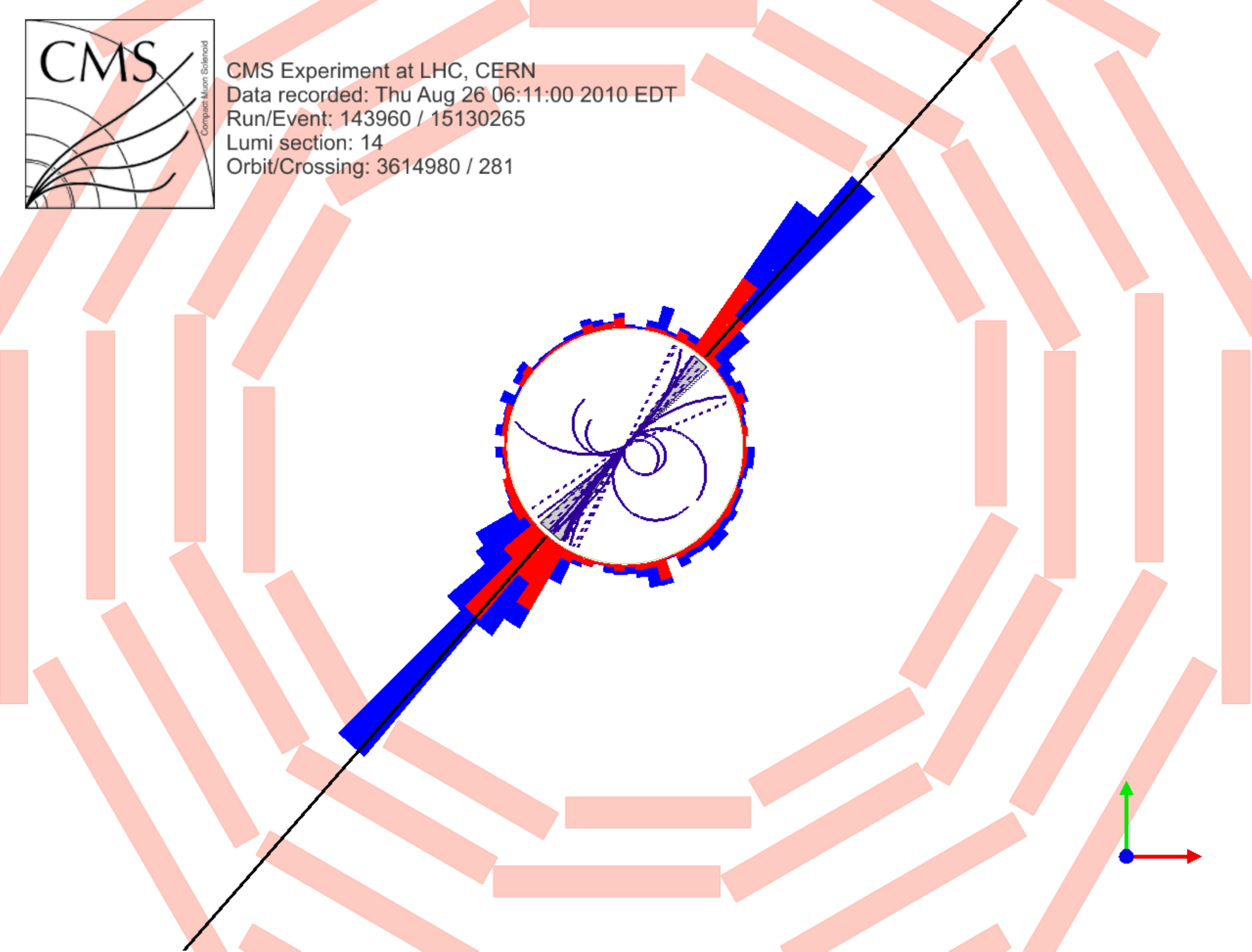}
\includegraphics[width=0.46\linewidth,bb=0 20 1400 1040,clip]{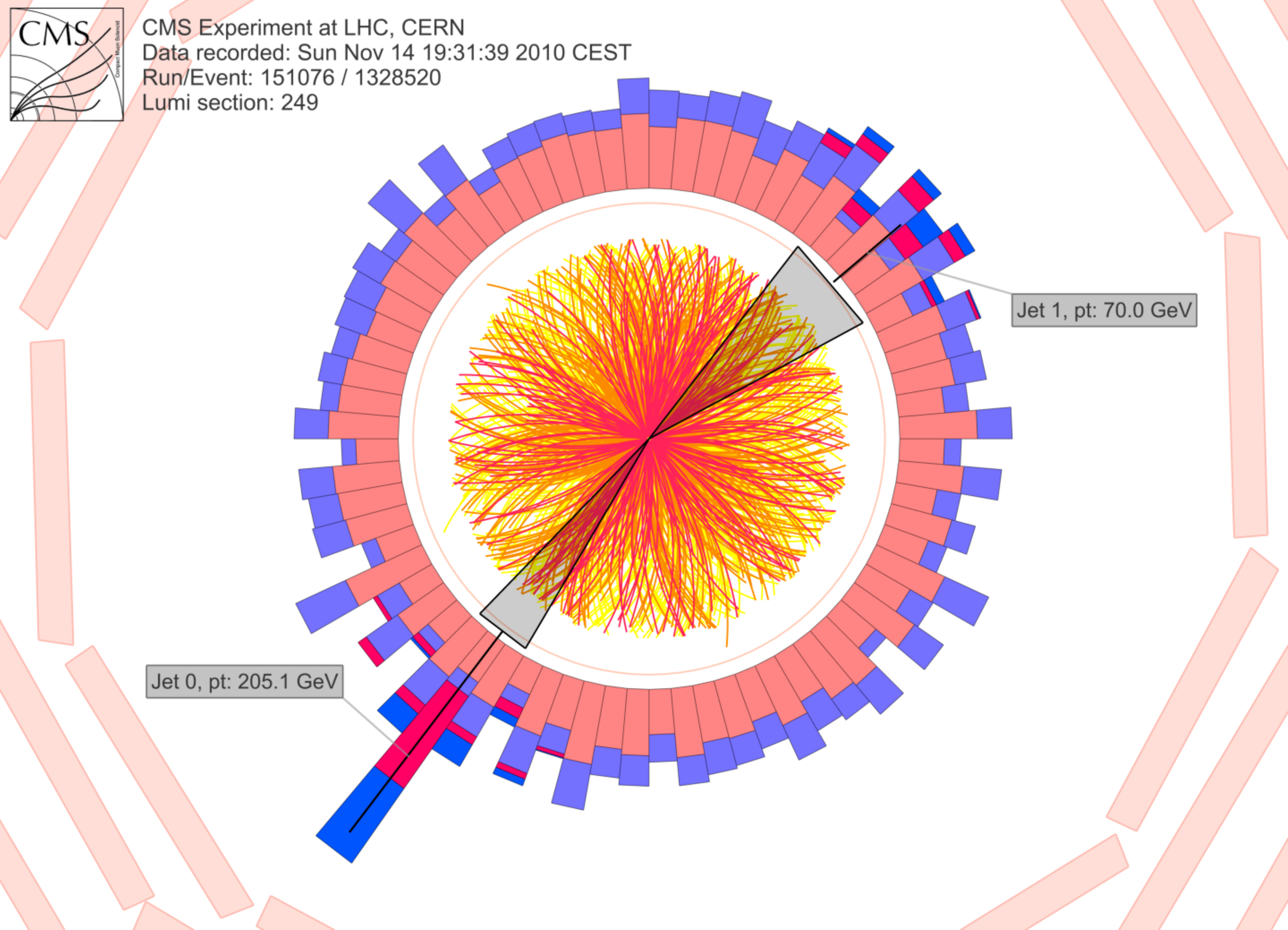}
\caption{Event display of dijet events in p+p (left) and Pb+Pb (right) collisions at $\sqrt{s_{\rm NN}}=2.76$~TeV from the CMS experiment~\cite{JetEventDisplay_CMS}.\label{fig:dijet}}
\end{center}
\end{figure}
\begin{figure}[hbt]
\begin{center}
\includegraphics[width=0.98\linewidth,bb=0 0 259 182]{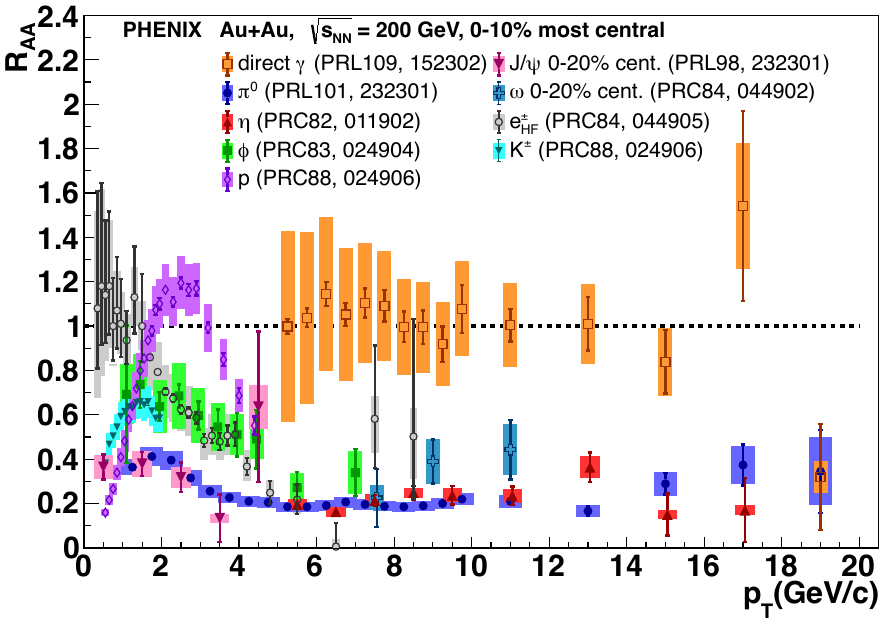}
\caption{Nuclear modification factor $R_{\rm AA}$ for various particles as a function of the transverse momentum in Au+Au collisions at $\sqrt{s_{\rm NN}}$ = 200 GeV from the PHENIX experiment. Figure was taken from Ref.~\cite{Sakaguchi:2019azf}\label{fig:RaaRHIC} and the references corresponding to each data are shown in the figure.}
\end{center}
\end{figure}
\begin{figure}[hbt]
\begin{center}
\includegraphics[width=0.98\linewidth,bb=0 0 567 427]{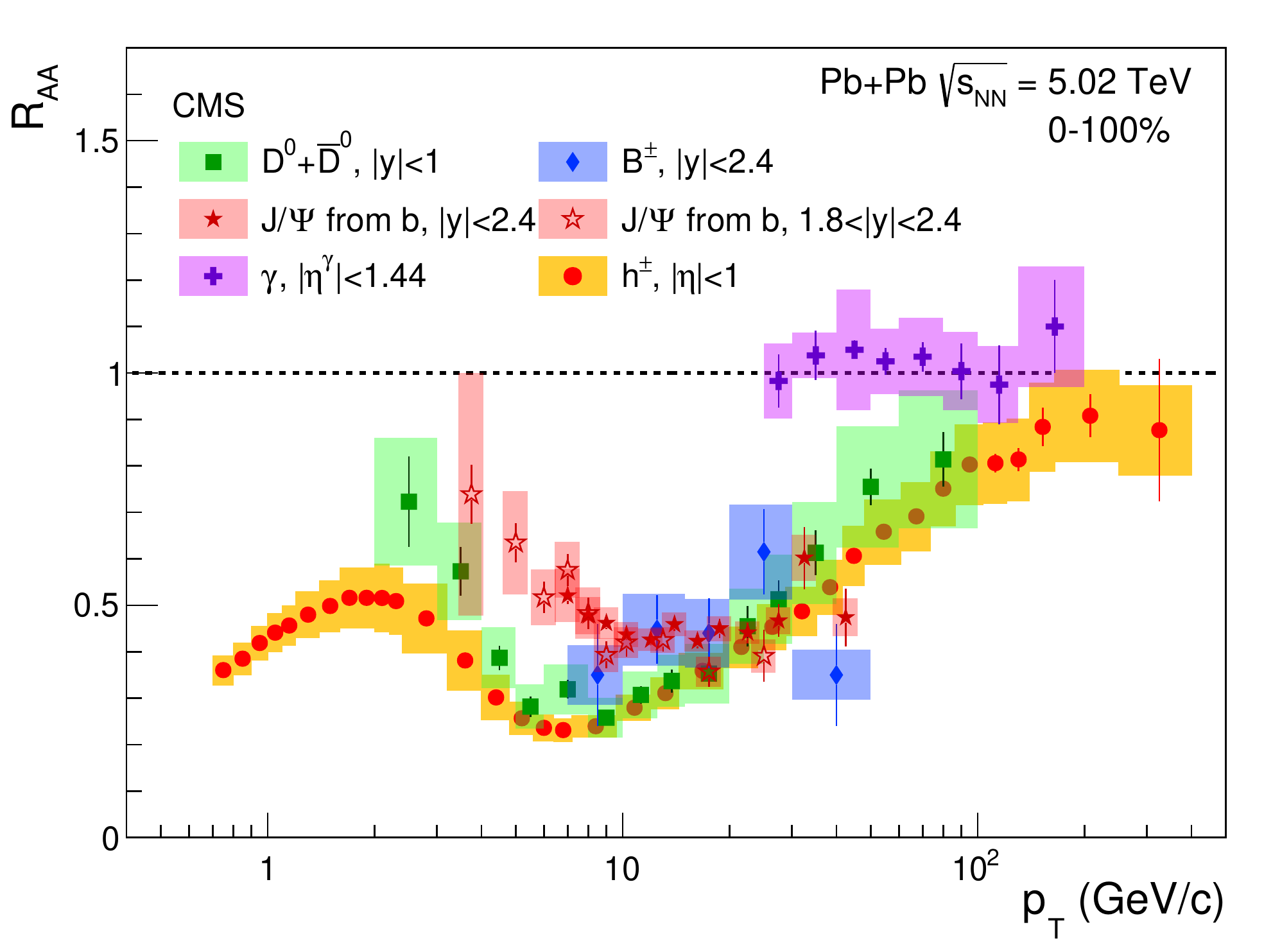}
\caption{Nuclear modification factor $R_{\rm AA}$ for various particles as a function of the transverse momentum in Pb+Pb collisions at $\sqrt{s_{\rm NN}}$ = 5.02 TeV from the CMS experiment~\cite{Sirunyan:2017xss,Sirunyan:2017isk,Sirunyan:2020ycu,Khachatryan:2016odn,Sirunyan:2017oug}.\label{fig:RaaLHC}}
\end{center}
\end{figure}

Figure~\ref{fig:dijet} shows event displays of jets in proton-proton (p+p) and nucleus-nucleus (A+A) collisions. Two collinear (``back-to-back") jets forced by momentum conservation are clearly observed in p+p collisions, while one of jets is significantly suppressed in A+A collisions: the phenomena called "jet quenching".
The jet quenching has been studied by measuring a so-called nuclear modification factor $R_{\rm AA}$ defined as the ratio of normalized single particle yields in p+p and A+A collisions:
\begin{equation}
    R_{AA} = \frac{d^2N_{\rm AA}/dp_{T}/dy}{\langle N_{\rm coll}\rangle d^2N_{\rm pp}/dp_{T}/dy},
\end{equation}
where $\langle N_{\rm coll}\rangle$ is the average number of binary nucleon-nucleon collisions. Figure~\ref{fig:RaaRHIC} shows $R_{\rm AA}$ of various particle species as a function of the transverse momentum $p_T$ measured by the PHENIX experiment. A strong suppression of light hadron production ($R_{\rm AA}<1$) at high $p_T$ was observed for the first time at RHIC, while $R_{\rm AA}$ of direct photons (photons produced at all stages through the system evolution except those from hadronic decays) is consistent with unity as expected since the photons do not interact via the strong force. The results on $R_{\rm AA}$ as well as on two-particle correlations~\cite{Adler:2002tq,Adams:2003im} show a significant energy loss of the partons in the hot medium which is not possible in ordinary nuclear matter, and therefore reveal a formation of quark-gluon plasma in heavy-ion collisions. 
Similar suppression in $R_{\rm AA}$ was also confirmed at the LHC with better precision for a wide range of kinematics~\cite{Aamodt:2010jd,CMS:2012aa} and was further investigated in heavy-flavour sector as shown in Fig.~\ref{fig:RaaLHC}. While the $R_{\rm AA}$ suppression for heavy flavour hadrons is quite similar to those for light hadrons at high $p_T$, there is a hint of mass-dependent radiative energy loss, i.e. $\Delta E_{u,d,s}>\Delta E_{c}>\Delta E_{b}$, in the low pt ($<$15 GeV/$c$) region.

These results in concert with theoretical models allow us to extract the medium properties such as the jet transport coefficient $\hat{q}$ characterized by an average transverse momentum transfer squared per unit length of the medium traversed. The detailed comparisons between the data and models determine $\hat{q}=1.2\pm0.3$ ${\rm GeV^2/fm}$ at RHIC and $1.9\pm0.7~{\rm GeV^2/fm}$ at the LHC for a quark with its energy  $E=10$~GeV~\cite{Burke:2013yra}. The extracted $\hat{q}$'s are two orders of magnitude larger than those for cold nuclear matter ($\sim$0.02~GeV$^{2}$/fm)~\cite{Ru:2019qvz}, supporting the finding that the extremely dense and opaque matter, the QGP, is created in the collisions. Recent studies from a Bayesian analysis~\cite{Soltz:2019aea} and lattice QCD calculation~\cite{Kumar:2020wvb} present the temperature dependence of $\hat{q}$ which agree with the previous work mentioned above.

\begin{figure}[hbt]
\begin{center}
\includegraphics[width=\linewidth,bb=10 10 833 466]{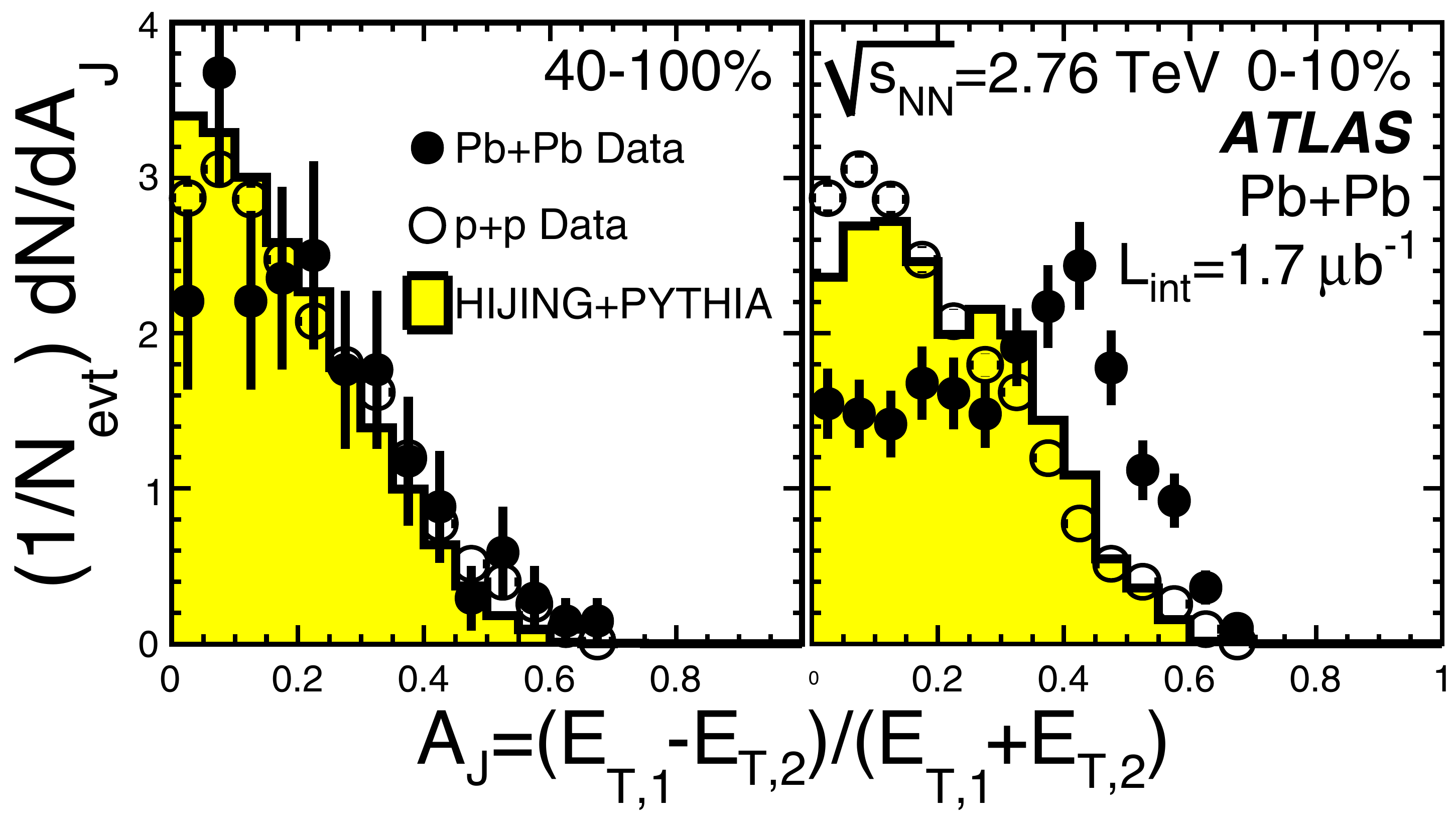}
\caption{Dijet asymmetry $A_J$ in the transverse energy for most central (right) and peripheral (left) events in Pb+Pb collisions as well as in p+p collisions at $\sqrt{s_{\rm NN}}=2.76$ TeV from the ATLAS experiment~\cite{Aad:2010bu}, compared with HIJING+PYTHIA calculations.\label{fig:Aj}}
\end{center}
\end{figure}

A jet can be reconstructed by final state particles using a clustering algorithm for a given jet radius and provides direct access to the initial partons and their energy loss. A momentum or energy imbalance of back-to-back jets was observed at the LHC as direct evidence of the jet quenching~\cite{Aad:2010bu,Chatrchyan:2012nia}, and later a similar trend was observed at RHIC~\cite{Adamczyk:2016fqm}.
Figure~\ref{fig:Aj} clearly shows an asymmetry in leading and subleading jet transverse energies for central A+A collisions, while less asymmetry is seen for peripheral collisions which is similar to what is seen in p+p collisions. The ``missing energy" of the jet seems to be redistributed to low momentum particles emitted to large angle relative to the jet momentum direction~\cite{Chatrchyan:2011sx,Khachatryan:2015lha} due to parton-medium interactions. More differential measurements such as jet fragmentation, jet substructure, and photon/$Z^0$-jet correlations have been started for better understanding the mechanism of the parton energy loss and parton-medium interaction, and to constrain the properties of the QGP (see recent review papers~\cite{Connors:2017ptx,Cao:2020wlm} for details).

%% file: 3-HydrodynamicalBehaviour.tex
\subsection{Anisotropic Flow}
In non-central collisions, the overlapped region of two Lorentz contracted nuclei is not circular but has an almond shape.  
The Emission pattern of the particles is influenced by the relation between the mean free path $\lambda$ of the particles and the size of the system $R$.  When $\lambda$ is larger than $R$, the particle emission is isotropic in the transverse direction. But, when $\lambda \ll R$, a hydrodynamic description is applicable and the particle emission becomes anisotropic.

Hydrodynamics has been considered to be applicable only to the system near the local equilibrium. However, it has been pointed out recently that this may not be true~\cite{Romatschke:2018prl}.
It is claimed that the criterion of the applicability of the hydrodynamics may be too strict and even in a system far from the local equilibrium hydrodynamic behavior may be seen. This may be related to hydrodynamic effects observed in small systems (Sec.~\ref{sec:small}).  




Hydrodynamic flow is derived by pressure gradients.  In the overlapped region of two nuclei, the pressure gradient is steeper in the plane of the reaction plane (plane defined by the impact parameter $\vec{b}$ and the beam axis), and because of that more particles are produced in plane than out-of plane. Thus the azimuthal distribution shows a characteristic $\cos (2\phi)$ modulation (elliptic flow), which is suggested to be important for the study of the hydrodynamic properties~\cite{Ollitraut:1992,Ollitraut:1993}. Thus, the initial spatial anisotropy of the almond shape is converted to the momentum anisotropy called the elliptic flow.  An important feature of the elliptic flow is that it is sensitive to the early stage of the collisions. Since the hot and dense region expands more in-plane, the spatial anisotropy disappears quickly as it expands. 


Experimentally, the azimuthal distribution is evaluated in terms of a Fourier expansion~\cite{Voloshin:1994mz},
\begin{equation}
E\frac{d^3N}{d^3p}=\frac{d^2N}{2\pi p_{_{\rm T}} d p_{_{\rm T}} dy} 
\left( 1+\sum_{n=1}^{\infty} 2 v_n \cos (n\phi) \right),
\label{eq;fourier}
\end{equation}
where $\phi$ is the azimuthal angle of produced particles with respect to the reaction plane.  The second-order coefficient $v_2$ quantifies the strength of the elliptic flow. 

\begin{figure}[hbt]
\begin{center}
\includegraphics[width=8cm]{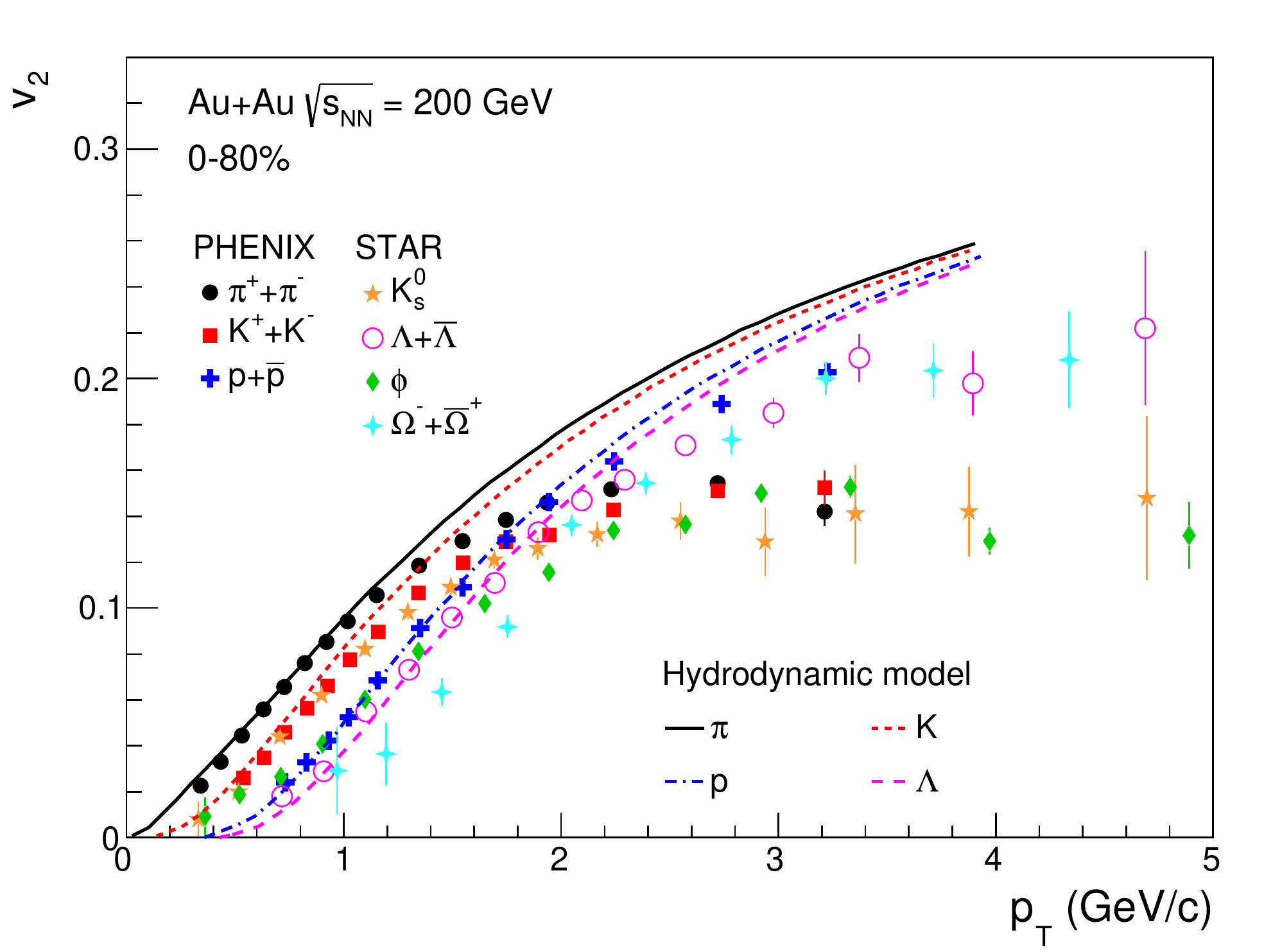}
\includegraphics[width=8cm,bb=0 0 567 477]{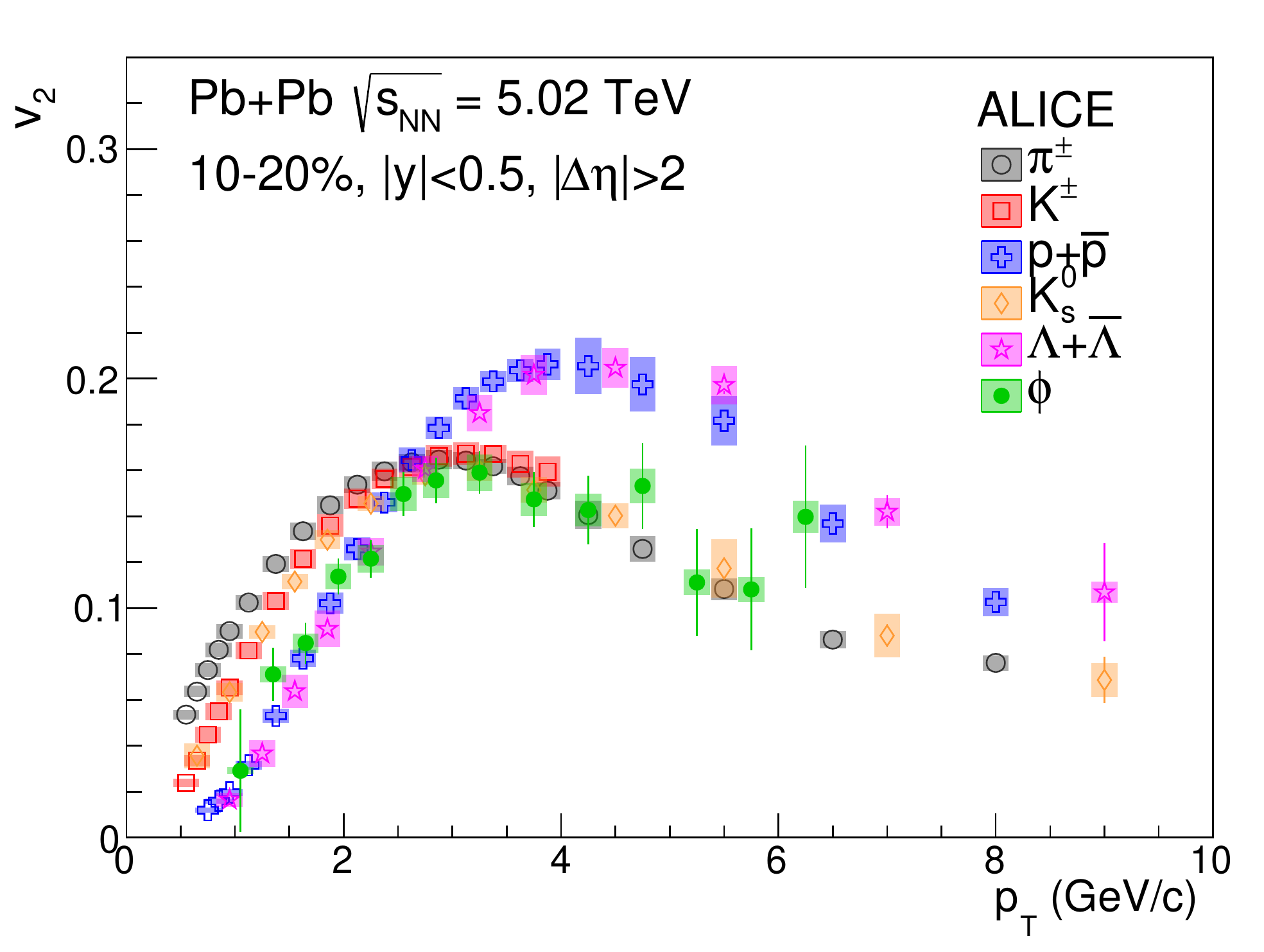}
\caption{(Upper panel) $v_2$ as a function of the transverse momentum for $\pi^{\pm}$, $K^{\pm}$, $K_s^0$, $p(\bar{p})$, $\phi$ $\Lambda(\bar{\Lambda})$, and $\Omega^-(\bar{\Omega}^+)$ in Au+Au collisions at $\sqrt{s_{\rm NN}}=200\ {\rm GeV}$ from the STAR and PHENIX experiments~\cite{Adams:2003am,Adamczyk:2015ukd,Adare:2006ti}. Solid and dashed curves show the prediction of the hydrodynamic model~\cite{Huovinen:2001cy}.  (Lower panel) $v_2$ as a function of the transverse momentum for $\pi^{\pm}$, $K^{\pm}$, $K_s^0$, $p(\bar{p})$, $\phi$, and $\Lambda(\bar{\Lambda})$ in semi-central Pb+Pb collisions at $\sqrt{s_{\rm NN}}=5.02\ {\rm TeV}$ from the ALICE experiment~\cite{ALICE-flow:2018}\label{fig:v2-rhic-alice}.}
\end{center}
\end{figure}

The top panel of Fig.~\ref{fig:v2-rhic-alice} shows elliptic flow $v_2$ for pions, kaons, protons, $\phi$, $\Lambda$, and $\Omega$ in mid-central Au+Au collisions at $\sqrt{s_{\rm NN}}=200\ {\rm GeV}$~\cite{Adams:2003am,Adamczyk:2015ukd,Adare:2006ti}.  In $p_{\rm T} < 2\ {\rm GeV/c}$, $v_2$ increases with $p_{\rm T}$ and a clear mass dependence is observed, which is well described by the hydrodynamic model as shown with solid and dashed curves~\cite{Huovinen:2001cy}.  On the other hand, in higher $p_{\rm T}$, there is a clear departure from the solid curves and two loci for mesons and baryons become visible. 
At the LHC, very similar behavior is observed: the bottom panel of Fig.~\ref{fig:v2-rhic-alice} shows $v_2$ for pions, kaons, protons, $\phi$, and $\Lambda$ in mid-central Pb+Pb collisions at $\sqrt{s_{\rm NN}}=5.02\ {\rm TeV}$~\cite{ALICE-flow:2018}.  As seen at RHIC, a mass dependence is seen in the low $p_{\rm T}$ region while grouping of mesons and baryons is observed also in higher $p_{\rm T}$, which suggests a different mechanism of particle production above $p_{\rm T} \gtrsim 2 $ GeV/c compared to the lower $p_T$ region, i.e. quark coalescence/recombination.

\subsection{Quark Coalescence/Recombination}
As a characteristic hadron production mechanism of the QGP, 
the quark coalescence/recombination picture has been introduced~\cite{Voloshin:2002wa,Molnar:2003ff,Fries:2003kq,Greco:2003xt,Hwa:2003ce}, in which quarks ($q$) and anti-quarks ($\bar{q}$) combine to mesons ($q \bar{q}$) and baryons ($qqq$).  This process becomes important at intermediate $p_{\rm T}$ region since production at high (low) $p_{\rm T}$ region is dominated by the fragmentation (thermal) process.   

To simplify the model, two assumptions are made; a)(anti-)quarks with the same momentum combine to form hadrons, b)(anti-)quarks have the universal elliptic flow $v_{2,q}(p_{\rm T})$.  Then the following relations are obtained:
\begin{eqnarray}
\frac{dN_{\rm M}}{d \phi}  &\propto (1+2 v_{2,q} \cos 2\phi)^2 \approx (1+4v_{2,q}\cos 2\phi),\\
\frac{dN_{\rm B}}{d \phi} &\propto (1+2 v_{2,q} \cos 2\phi)^3 \approx (1+6v_{2,q}\cos 2\phi),
\end{eqnarray}
where $N_{\rm M}$ and $N_{\rm B}$ are yields of the meson and the baryon. 
Thus, the elliptic flow for mesons ($v_{2,{\rm M}}$) and baryons ($v_{2,{\rm B}}$) are scaled according to the number of constituent quarks $n_q$ (quark number scaling) as,
\begin{eqnarray}
v_{2,{\rm M}}(p_{\rm T}) \sim 2 v_{2,q}(p_{\rm T}/2), \ v_{2,{\rm B}}(p_{\rm T}) \sim 3 v_{2,q}(p_{\rm T}/3).
\end{eqnarray}

\begin{figure}[hbt]
\begin{center}
\includegraphics[width=8.6cm]{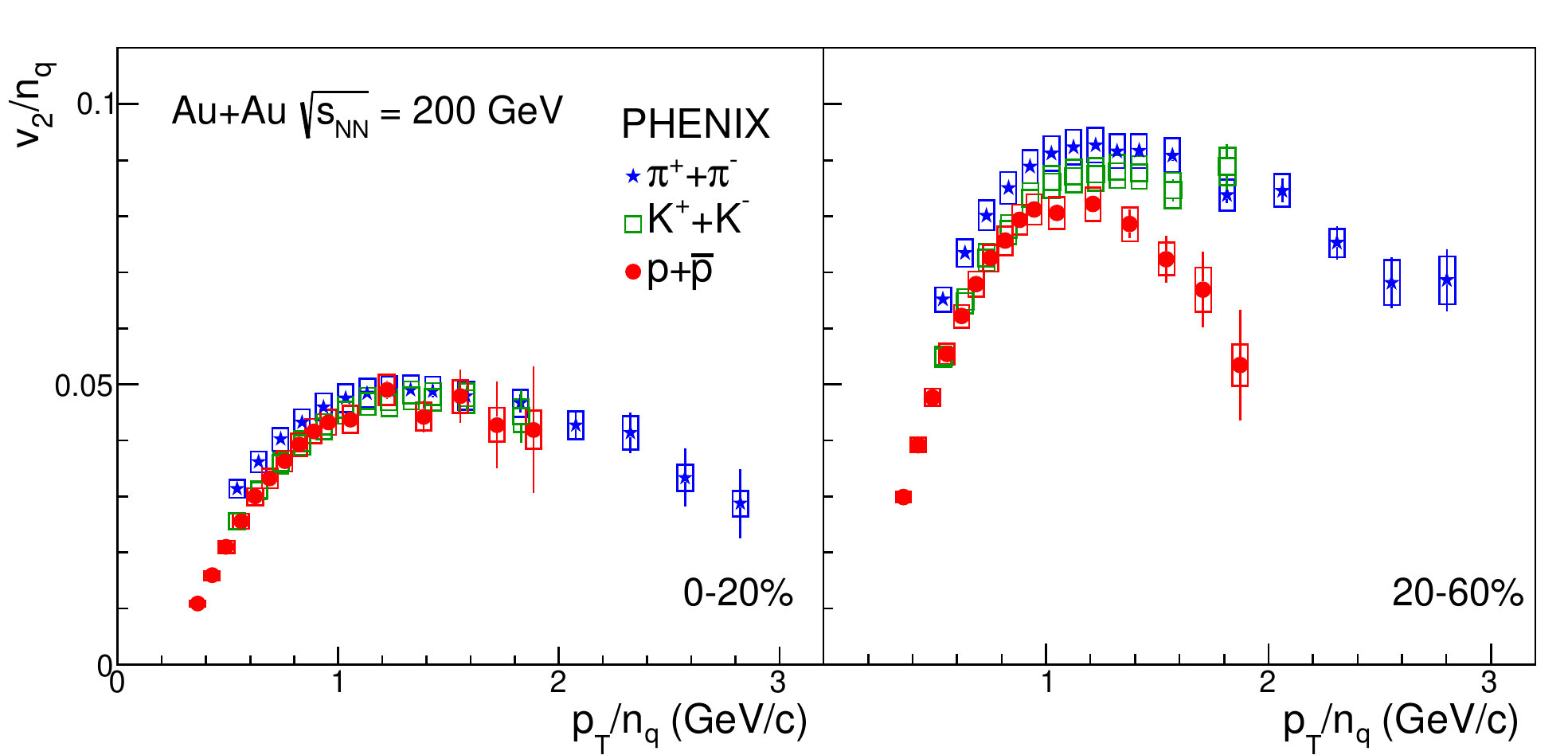}
\caption{$v_2/n_q$ of $\pi^{\pm}$, $K^{\pm}$, and $p(\bar{p})$ as a function of the transverse momentum per quark, $p_{\rm T}/n_q$ in central (left) and mid-central (right) Au+Au collisions at $\sqrt{s_{\rm NN}}=200{\rm \ GeV}$ from the PHENIX experiment.~\cite{PHENIX-QNS:2012}.
\label{fig:v2-QNS-PHENIX}}
\end{center}
\end{figure}

In Fig.~\ref{fig:v2-QNS-PHENIX}, $v_2/n_q$ as a function of transverse momentum per quark, $p_{\rm T}/n_q$, in central ($0-20\%$ centrality) and mid-central ($20-60\%$ centrality) Au+Au collisions at $\sqrt{s_{\rm NN}}=200{\rm \ GeV}$ are shown. In central collisions, $v_2/n_q$ of pions, kaons, and protons agree with each others within the statistical and systematic uncertainties, which supports the quark coalescence picture.  But, in  peripheral collisions, a clear departure from the quark number scaling is observed for $p_{\rm T}>1.3{\rm \ GeV/c}$~\cite{PHENIX-QNS:2012}.  
It is expected that the scaling does not work at high $p_{\rm T}$ region, where the fragmentation process becomes dominant.
At the LHC, the scaling has been observed approximately at the level of $\pm 20\%$~\cite{ALICE-flow:2018}. 

\subsection{Higher Order Harmonics}

\begin{figure}[hbt]
\begin{center}
\includegraphics[width=8cm,bb=0 0 1107 667]{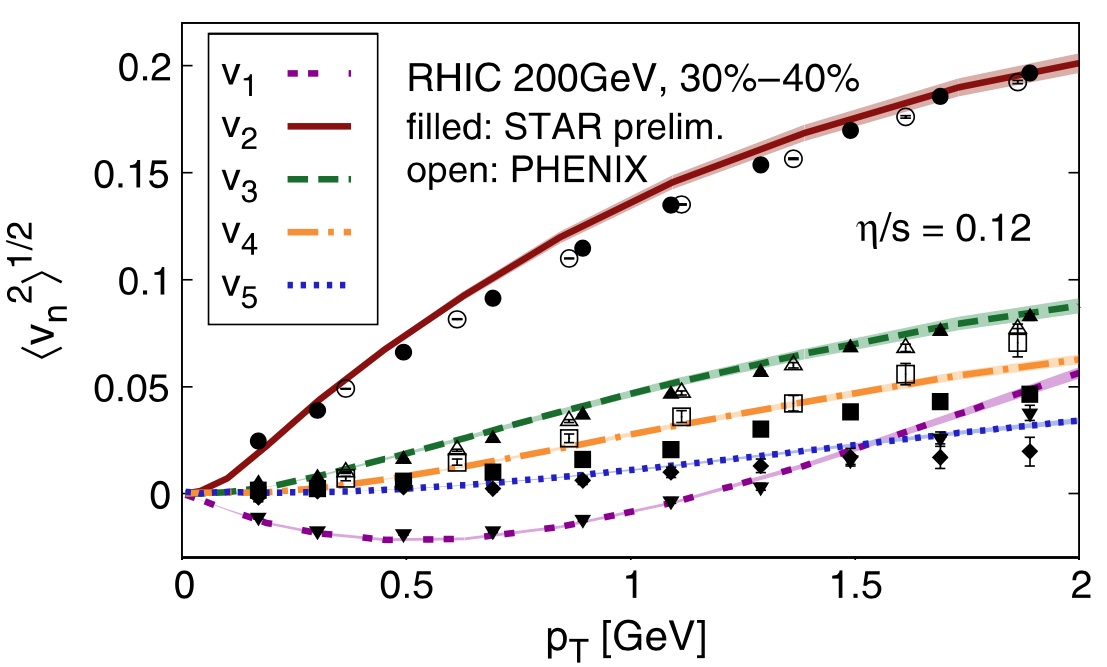}
\caption{Root-mean-square anisotropic flow coefficients $\langle v^2_n \rangle^{1/2}$ as a function of the transverse momentum in mid central Au+Au collisions at $\sqrt{s_{\rm NN}}=200\ {\rm GeV}$.  Curves show hydrodynamic model calculations~\cite{Gale:2012rq} assuming the shear viscosity $\eta/s = 0.12$\label{fig:v345-rhic}. Experimental data are from PHENIX~\cite{Adare:2011tg} and STAR~\cite{Pandit:2012mq} experiments. }
\end{center}
\end{figure}

Not only the second Fourier coefficient, the elliptic flow, but also higher order Fourier coefficients have been measured.  While the elliptic flow $v_2$ arises from the almond shape of the initial overlapped region, higher order harmonics are primarily due to the initial fluctuations of the geometry. In other words, because of the limited number of nucleons involved in the collisions, there are event-by-event fluctuations in the nucleon position and the distribution.  Such geometrical fluctuations are converted through the hydrodynamic expansion and observed as the higher flow harmonics. Figure~\ref{fig:v345-rhic} shows observed $v_n$, $n=$1--5, compared with the hydrodynamic model calculations~\cite{Gale:2012rq}.  The experimental data are from PHENIX~\cite{Adare:2011tg} and STAR~\cite{Pandit:2012mq} collaborations.

The hydrodynamic calculations start from the equilibrium state of the QGP after a very short ($<1{\rm \ fm}/c$) pre-equilibrium state and compute the  expansion, in which shear viscosity is included as $\eta/s$ (the ratio of shear viscosity $\eta$ to the entropy density $s$), followed by hadronic expansions.  As seen in Fig.~\ref{fig:v345-rhic}, an agreement between the experiment and the theory is striking: the model with the intrinsic fluctuations reproduces the higher flow harmonics as well as the elliptic flow assuming the shear viscosity of $\eta/s = 0.12$, which is very close to the theoretical lower limit of $\frac{1}{4\pi}$~\cite{Kovtun:2004de}.

\begin{figure}[hbt]
\begin{center}
\includegraphics[width=0.49\linewidth,bb=0 0 567 534]{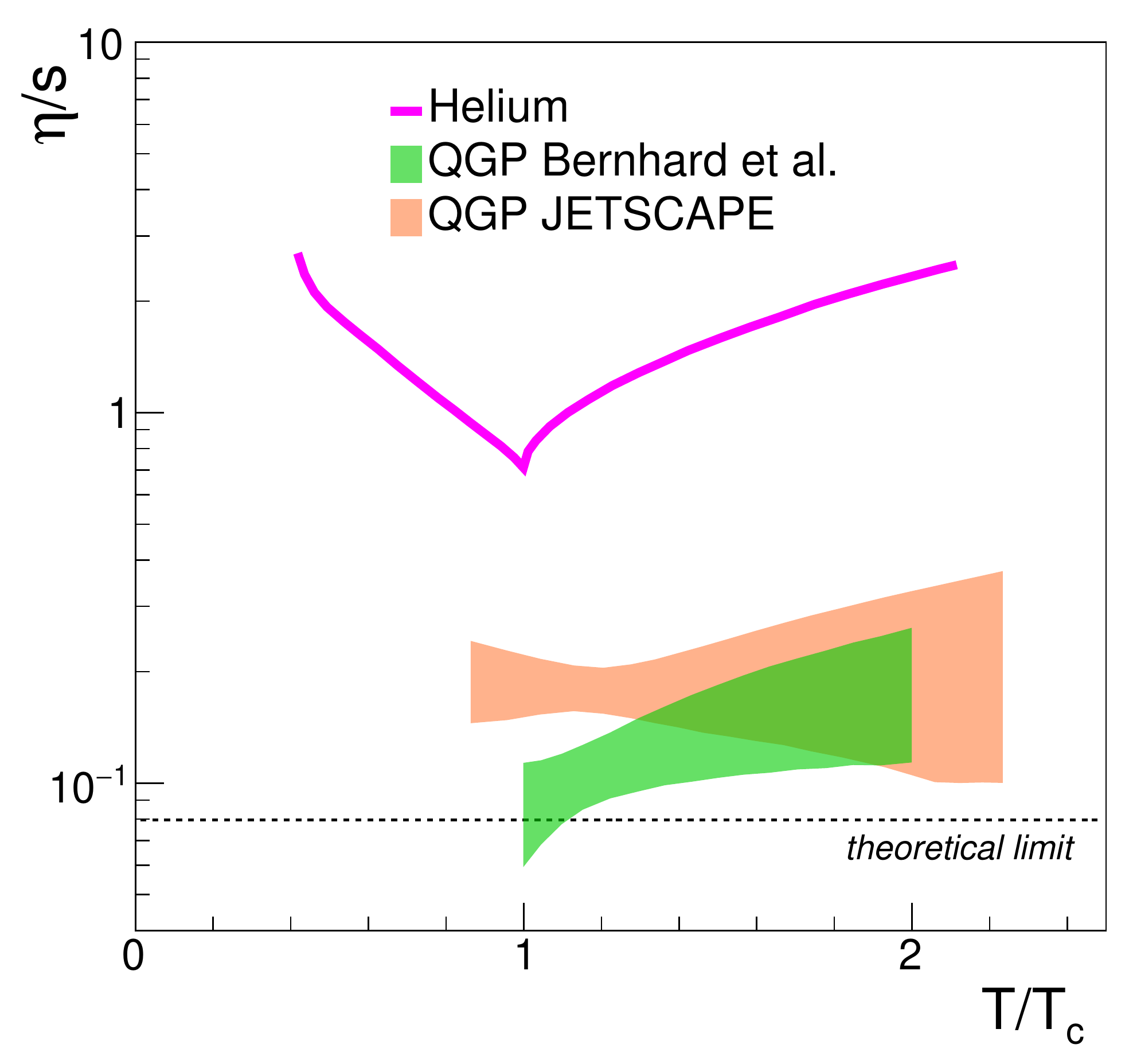}
\includegraphics[width=0.49\linewidth,bb=0 0 567 534]{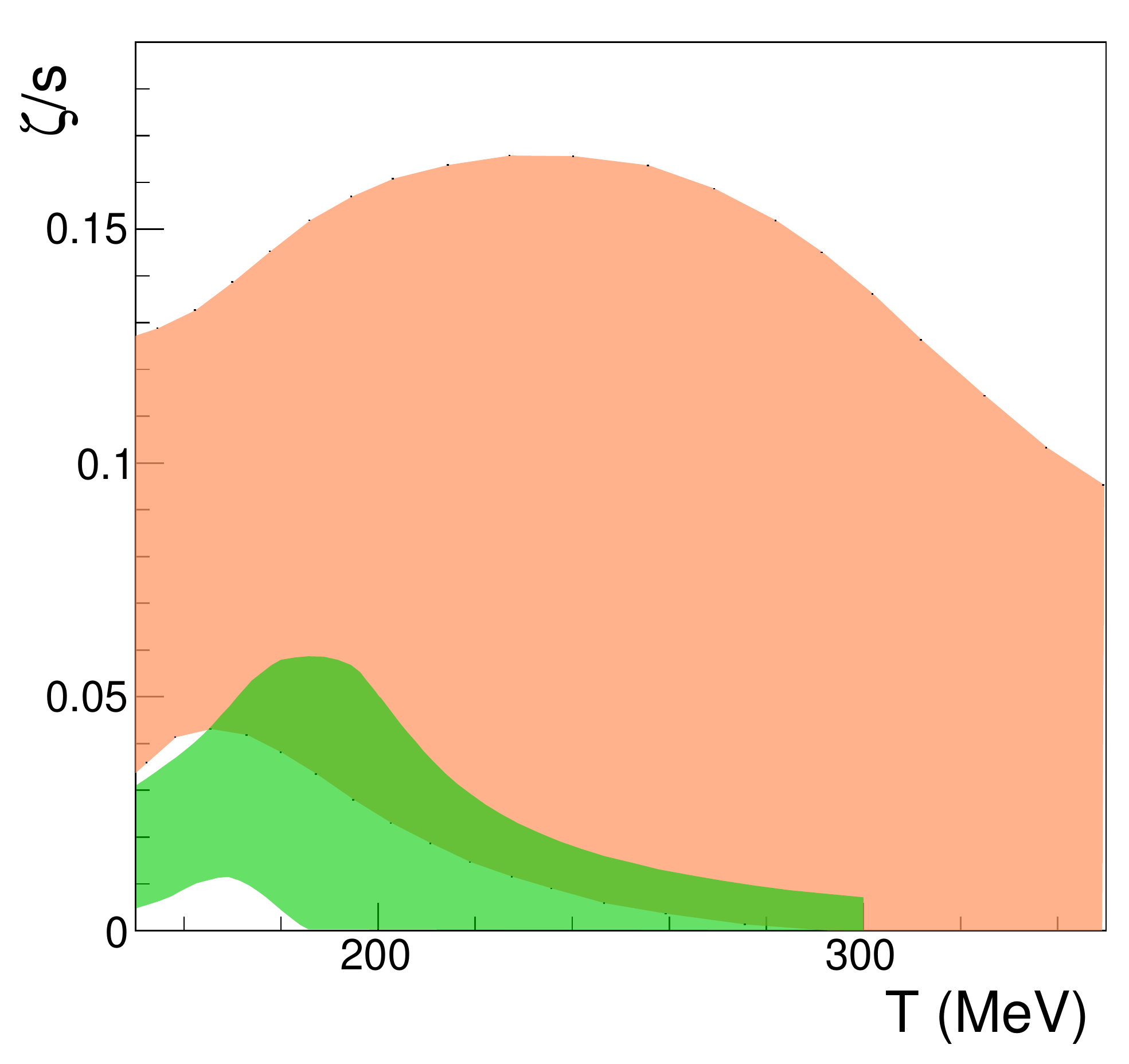}
\caption{Estimated temperature-dependent specific shear (left) and bulk (right) viscosities of the QGP using Bayesian method~\cite{Bernhard:2019bmu,Everett:2020yty}.  The shaded bands show $90 \%$ credible region for the QGP $\eta/s(T)$ and $\zeta/s$ estimated from experimental data. The green bands are from Ref.~\cite{Bernhard:2019bmu} and the orange bands from Ref.~\cite{Everett:2020yty}. The pseudo-critical temperature $T_c=156.5\pm1.5$ MeV~\cite{Bazavov:2018mes} for a crossover phase transition from the QGP to hadronic matter is assumed. Solid line shows $\eta/s(T)$ for helium relative to its critical temperature and dashed line for theoretical lower limit~\cite{Kovtun:2004de}.
\label{fig:viscosity}}
\end{center}
\end{figure}

In order to extract the properties of the QGP and constrain the initial conditions, a state-of-the-art calculation, Bayesian parameter estimation methods have been applied and the temperature-dependent specific shear and bulk viscosity have been extracted.  
Shear viscosity is known to primarily affect the collective behavior and the azimuthal anisotropy, while bulk viscosity also affects the collective behavior in particular radial flow and mean $p_{\rm T}$ of hadrons. To evaluate these key physics properties with quantitative uncertainties, this method has been applied using many experimental observables at the same time.
Results of two independent studies with this method are shown in Fig.~\ref{fig:viscosity}.  
In ref.~\cite{Bernhard:2019bmu},
parameters in the hydrodynamic model are carefully studied to constrain the range of each parameter according to the experimental data such as yields of charged particles, transverse energy, yields of pions, kaons and protons as well as mean $p_{\rm T}$ of pions, kaons, protons, and azimuthal anisotropies ($v_n$, $n=$1--4) in Pb-Pb collisions at $\sqrt{s_{\rm NN}}$ = $2.76{\rm \ and \ }5.02{\rm \ TeV}$.  
In ref.~\cite{Everett:2020yty},
both RHIC and the LHC data are used. 
Parameter constraining methods and conditions are different in these studies, which also leads to slightly different results of the shear and bulk viscosities and their uncertainties.
In Fig.~\ref{fig:viscosity}, the shear viscosity $\eta/s$ and the bulk viscosity $\zeta/s$ are shown as a function of temperature $T$.  The shear viscosity is compared with that of  helium at its critical pressure. 
As seen in the figure, the extracted $\eta/s$ of the QGP is much smaller than that of helium, showing that the QGP is a nearly non-viscous fluid.  The right panel of Fig.~\ref{fig:viscosity} shows the bulk viscosity $\zeta/s$ as a function of temperature.
In the early days, hydrodynamic calculations used to assume that the bulk viscosity is negligible.  But, these studies have successfully provided the most reliable constraints on the shear viscosity as well as the bulk viscosity, providing a better description of the experimental data.  

While it is not clear due to the large uncertainty, $\eta/s$ tends to increase at higher temperature in
Fig.~\ref{fig:viscosity}. 
At RHIC and the LHC, the high opacity and the non-viscous fluid nature, i.e. short mean free path/large cross section in the fluid, suggest a strongly-coupled QGP.
On the other hand, in much higher energy collisions like at CERN-FCC\cite{CERN-FCC}, the viscosity may increase leading to a more viscous fluid, and we may have a chance to observe even a weakly-coupled QGP instead of the strongly-coupled QGP.

Many hydrodynamic model calculations have been carried out and they have shown two important features; very low $\eta/s$ and very short pre-equilibrium state before the QGP established ($t<1{\ \rm fm/c}$).  Mechanism of the rapid thermalization is not understood yet, where one missing piece of the information is the initial stage of the collisions.  For this information, measurements of small-$x$ gluon distribution via direct photon and jets are proposed at the LHC~\cite{FOCAL_LOI}.

%% file: 4-OtherSignaturesIssues.tex

\subsection{Direct photons and their puzzle}
\begin{figure}[t]
\begin{center}
\includegraphics[width=8cm,bb=0 0 600 518]{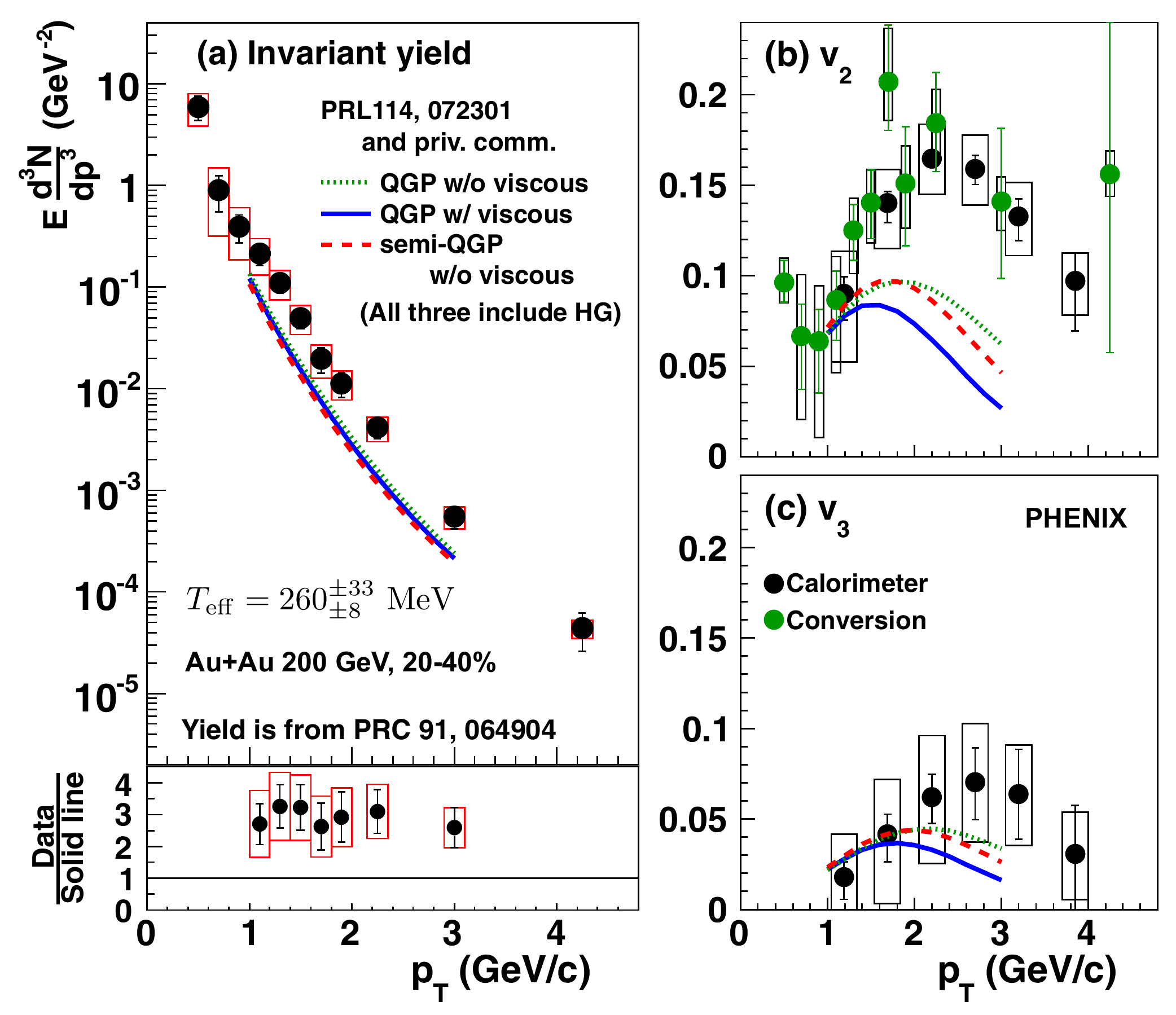}
\caption{Direct photon yield (a) and elliptic (b) and triangular (c) flow as a function of the transverse momentum~\cite{Adare:2015lcd} comparing to hydrodynamic model calculations with different assumptions. An effective temperature $T_{\rm eff}$ extracted from the inverse slope of the yield is shown in the figure. See Ref.~\cite{Adare:2015lcd} and references therein for details. \label{fig:phenix_photon}}
\end{center}
\end{figure}

Photons are a penetrating probe because of their small cross section and can be used to study the space-time evolution of the system since they are produced at all stages through the collision. Transverse momentum distributions of direct photons have been measured at both RHIC and the LHC~\cite{Adare:2008ab,Adare:2014fwh,Adam:2015lda}, where the enhancement at low $p_T$ ($<4$~GeV/$c$) in central A+A collisions relative to the scaled p+p data is described well by perturbative QCD. The excess indicates the photon production due to thermal radiation from the QGP.  Figure~\ref{fig:phenix_photon}(a) shows the direct photon yields after the subtraction of scaled p+p yield. An effective temperature extracted from the excess is found to be $T_{\rm eff}=260^{\pm 33}_{\pm 8}$~MeV for 20-40\% Au+Au collisions at RHIC top energy and $T_{\rm eff}=297\pm12\pm41$~MeV for 0-20\% Pb+Pb collisions at the LHC, both of which are much hotter than the critical temperature $T_{c}$ discussed in Sec.~\ref{sec:PT}. 

Since photons are predominantly emitted at early times (high temperature), one expects that azimuthal anisotropy of direct photons would be small because the flow is developed later in time with the collective expansion of the system as discussed in Sec.~\ref{sec:hydro}. However, experimental data at RHIC and the LHC show a sizable $v_2$ (even $v_3$) of direct photons comparable to the hadron $v_2$~\cite{Adare:2011zr,Adare:2015lcd,Acharya:2018bdy}. Figure~\ref{fig:phenix_photon} shows the direct photon yield and $v_2$ and $v_3$ measured at RHIC, comparing to hydrodynamic calculations. So far none of models can satisfactorily explain both the yield and $v_2$ simultaneously, which is known as ``photon puzzle"  (see Ref.~\cite{David:2019wpt} for recent review). There are still ongoing efforts from the experimental side to reduce the uncertainty as well as ones from the theoretical side with new ideas.

\subsection{Debye screening effect}
Suppression of heavy quarkonium has been proposed as a signature of the QGP~\cite{Matsui1986}: heavy quarkonium such as charmonium or bottomonium is expected to dissolve in the QGP when the potential between the quarks is screened by copious color charges of quark and gluon in the plasma, i.e. the Debye screening effect.  

Since the first measurement of the $J/\Psi$ yields in Pb+Pb collisions at $\sqrt{s_{\rm NN}}=17$ GeV at SPS~\cite{Abreu:1999qw}, heavy quarkonia have been measured rigorously at RHIC and also at the LHC. Quantitative understanding of the yields at higher energies is found to be complicated since there are at least two competing effects: suppression due to the screening effects and the enhancement due to the recombination process.   

\begin{figure}[hbt]
\begin{center}
\includegraphics[width=8cm,bb=0 0 853 778]{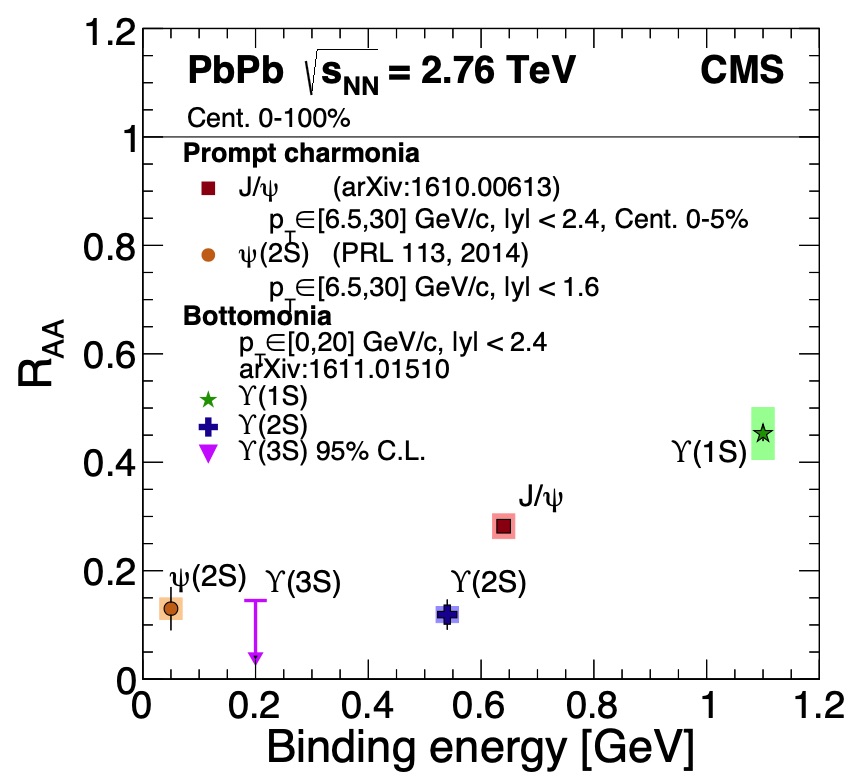}
\caption{$R_{\rm AA}$ of quarkonium states as a function of the binding energy in Pb+Pb collisions at $\sqrt{s_{NN}}=2.76$ TeV measured by the CMS experiment, which is consistent with the sequential suppression pattern~\cite{Chatrchyan_2012,Rev_bottomonium_CMS,Fate_of_quarkonia}. \label{fig:CMS_Seq_Suppr}}
\end{center}
\end{figure}

Quarkonia larger than the Debye length, the range of the interaction, are dissolved in the plasma.
Therefore, the weaker bound quarkonium (larger radius) is expected to dissolve more completely compared to the stronger bound quarkonium (smaller radius). Thus, when suppressions of various quarkonia are compared, sequential suppression of their yields is expected according to their radii/binding energies.  
As shown in Fig.\ref{fig:CMS_Seq_Suppr}, $R_{\rm AA}$ of charmonia, $J/\psi$ and $\psi$(2S), and bottomonia, $\Upsilon$(1S) and $\Upsilon$(2S), have been measured and the sequential suppression behavior has been clearly observed at the CMS experiment~\cite{Chatrchyan_2012,Rev_bottomonium_CMS,Fate_of_quarkonia} which supports the assumption of QGP formation. 
For quantitative understanding, further theoretical and experimental studies are needed.

\subsection{Electromagnetic field/Chirality}
In heavy-ion collisions, a short-lived strong magnetic field is created due to the moving electric charges carried by protons inside (fragmented)nuclei. The peak magnitude is expected to reach $10^{13{\mathchar`-}14}$~Tesla~\cite{Kharzeev:2007jp,Skokov:2009qp}, stronger by a few orders than the magnetic field on the surface of neutron star called magnetars. 
The presence of such a strong magnetic field has not been confirmed experimentally and the measurement of the strong field itself is of great interest. The lifetime of the field has large uncertainty and could be significantly extended depending on the electric conductivity of QGP due to Faraday's law of induction~\cite{Tuchin:2013apa,McLerran:2013hla,Skokov:2009qp}. In other words, one can probe the conductivity of QGP by studying the QGP response to the strong magnetic field generated by the charged spectator fragments.
It is suggested that the effect of the magnetic field appears in difference of directed flow (the first-order coefficient in Eq.~\ref{eq;fourier}) between particles and antiparticles~\cite{Rybicki:2013qla,Gursoy:2014aka}, although the uncertainty is still too large to make any statement~\cite{Acharya:2019ijj,Adam:2019wnk}. 

Not only the magnetic field but also the electric field should be created in the initial state particularly for asymmetric collisions, e.g. Cu+Au. The effect appears in the charge difference of directed flow due to the Coulomb force, which is sensitive to the electric conductivity of QGP~\cite{Hirono:2012rt} and the time evolution of charge creation, i.e. quark and antiquark production~\cite{Voronyuk:2014rna}. Experimental result shows such a charge difference in hadron directed flow~\cite{Adamczyk:2016eux}, indicating the presence of the initial electric field. Comparing to theoretical model with the electric field~\cite{Voronyuk:2014rna}, only $\sim$10\% of all (anti)quarks produced in the collisions are found to be created at that time when the electric field is strong ($t<0.5$~fm/$c$).

Chiral symmetry is spontaneously broken in the QCD vacuum but under high temperature and/or high density the chiral symmetry is restored where chirality is well defined. It is proposed that the presence of the initial strong magnetic field with QGP leads to chiral phenomena such as the chiral magnetic effect (CME): the phenomenon that electric current is induced along the magnetic field under chirality imbalance created by topological fluctuations of QCD vacuum~\cite{Fukushima:2008xe,Kharzeev:2007jp}. Such an electric current, i.e. charge separation of produced particles, has been extensively studied via two-particle correlations relative to the reaction plane~\cite{Voloshin:2004vk} at RHIC and the LHC, however the definitive conclusion is not yet reached because of significant contributions from backgrounds~\cite{Kharzeev:2015znc}. Analysis of isobar collision ($^{96}_{44}$Ru+$^{96}_{44}$Ru and $^{96}_{40}$Zr+$^{96}_{40}$Zr) data is ongoing at the STAR experiment. The two species have the same mass number (similar background) but different electric charges ($>$10\% difference in the magnetic field), therefore it is expected that the measurements provide a definitive answer for CME.

\subsection{Vorticity and polarization}
\begin{figure}[hbt]
\begin{center}
\includegraphics[width=8cm,bb=0 0 567 450]{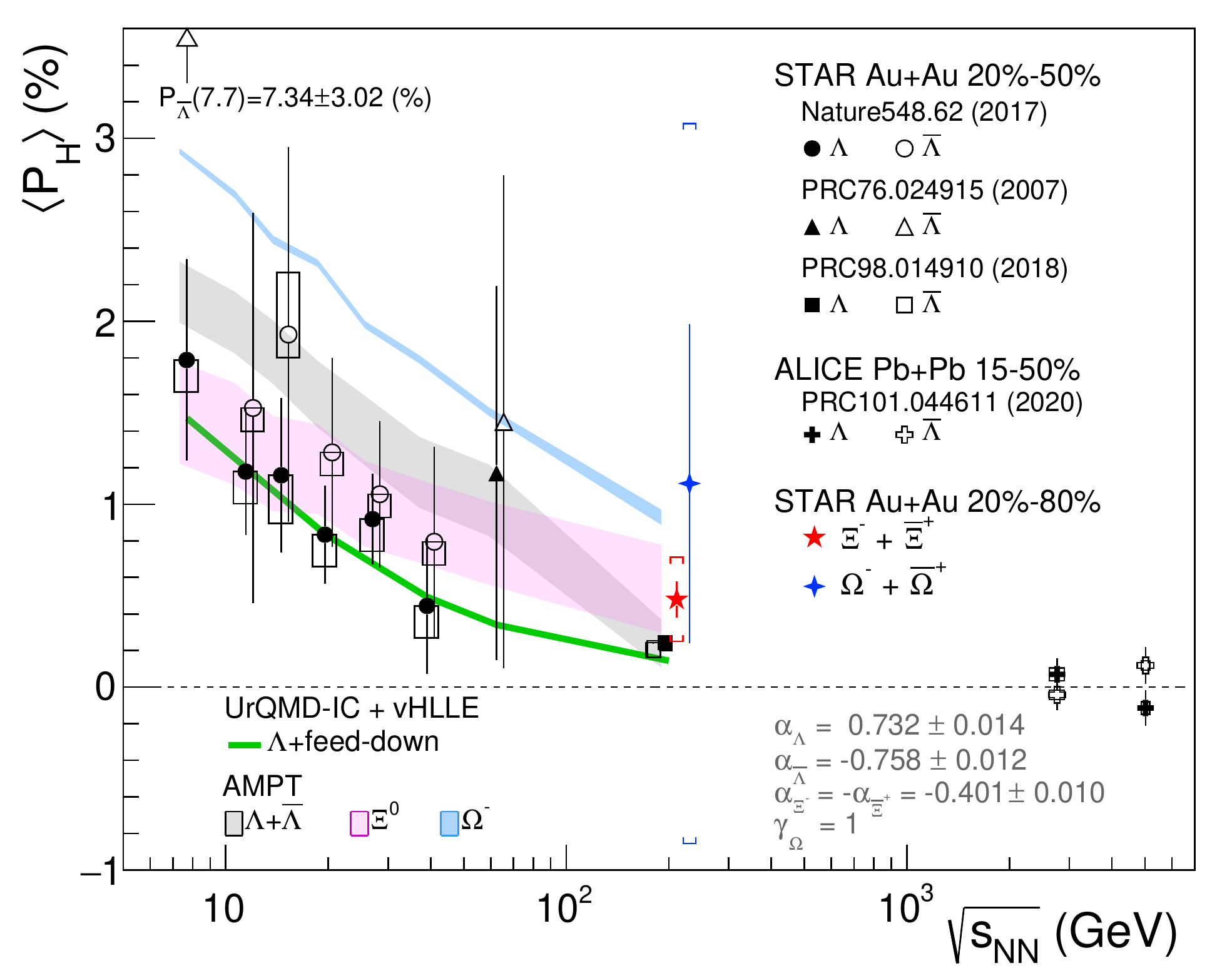}
\caption{Global polarization as a function of the collision energy for $\Lambda$ and $\bar{\Lambda}$ hyperons as well as for $\Xi$ and $\Omega$ hyperons~\cite{Abelev:2007zk,STAR:2017ckg,Adam:2018ivw,Acharya:2019ryw,Adam:2020pti}. Solid line and shaded bands denote calculations from viscous hydrodynamic model (vHLLE) with UrQMD initial condition~\cite{Karpenko:2016jyx} as well as a multi-phase transport (AMPT) model~\cite{Wei:2018zfb} respectively. Decay parameters $\alpha$ and $\gamma$ of each particles~\cite{Zyla:2020zbs} used in the measurements are shown in the figure. Note that the old results are corrected for recent update of the decay parameter.\label{fig:PH}}
\end{center}
\end{figure}

Similar to the initial magnetic field, large orbital angular momentum is expected to be created in the initial state for non-central collisions. A fraction of the orbital angular momentum would be transferred to the created matter, leading to global polarization of produced particles due to spin-orbit coupling~\cite{Liang:2004ph,Voloshin:2004ha,Becattini:2007sr}. The ``global" means net spin alignment along the initial angular momentum direction which is perpendicular to the reaction plane and coincides with the direction of the initial magnetic field. Global polarization of $\Lambda$ and $\bar{\Lambda}$ hyperons was measured at the STAR experiment and the extracted vorticity is found to be  $\omega$$\sim$$10^{22}~s^{-1}$~\cite{STAR:2017ckg,Adam:2018ivw}. The matter created in the collisions is realized as the most vortical fluid ever observed. The polarization may also help to constrain the lifetime of the initial magnetic field since the polarization due to magnetic-spin coupling differs in the sign between particles and antiparticles.  Theoretical models~\cite{Karpenko:2016jyx,Wei:2018zfb,Sun:2017xhx,Xie:2017upb,Li:2017slc} can describe the energy dependence of the polarization quantitatively as shown in Fig.~\ref{fig:PH} where calculations from viscous hydrodynamic model~\cite{Karpenko:2016jyx} and a multi-phase transport model~\cite{Wei:2018zfb} are compared, for both of which local thermal equilibrium is assumed and the polarization is calculated based on thermal vorticity at freeze-out~\cite{Becattini:2007sr,Becattini:2013fla}.
On the other hand, discrepancies between the data and models are seen in differential measurements and those issues need to be resolved~\cite{Niida:2018hfw,Adam:2019srw,Becattini:2020ngo}. For better understanding the nature of vorticity and spin dynamics in heavy-ion collisions, the measurement has been recently extended to other hyperons, $\Xi$ and $\Omega$~\cite{Adam:2020pti}, as shown in Fig.~\ref{fig:PH}. These measurements will provide new information on the dynamics of the QGP and open new directions to study QCD matter under extremely strong magnetic and vorticity fields~\cite{Jiang:2016wvv,Fujimoto:2021xix}.

\subsection{QGP droplet in small system?}\label{sec:small}
\begin{figure}[hbt]
\begin{center}
\includegraphics[width=0.8\linewidth,bb=0 0 856 737]{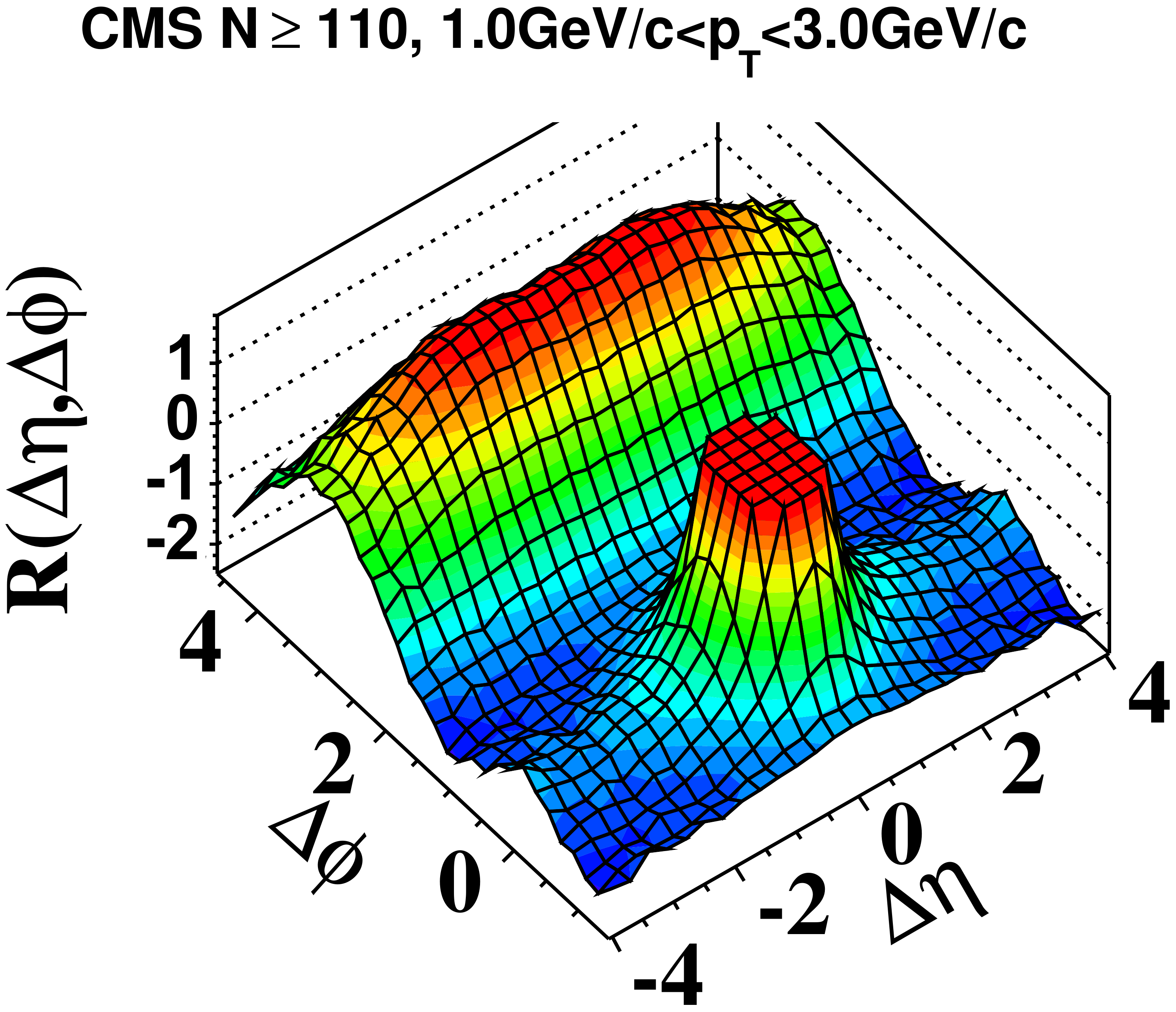}
\caption{Two-particle correlation as functions of relative azimuthal angle $\Delta\phi$ and pseudorapidity $\Delta\eta$ for high multiplicity events in p+p collisions at $\sqrt{s_{\rm NN}}=7$~TeV from the CMS experiment~\cite{Khachatryan:2010gv}.\label{fig:CMS_ppRidge}}
\end{center}
\end{figure}

\begin{figure}[hbt]
\begin{center}
\includegraphics[width=\linewidth,bb=5 0 358 121]{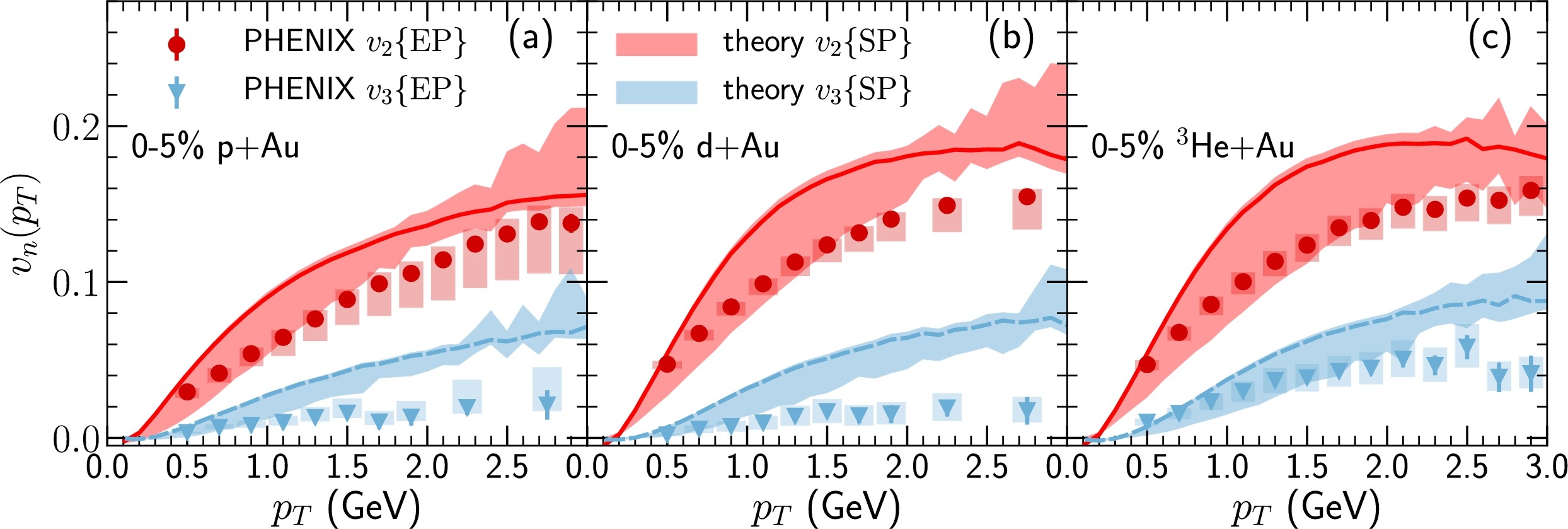}
\caption{Elliptic and triangular flow coefficients $v_n$ of charged hadrons as a function of the transverse momentum in p+Au, d+Au, and $^{3}$He+Au collisions at $\sqrt{s_{\rm NN}}$ = 200 GeV from the PHENIX experiment~\cite{PHENIX:2018lia}. Solid lines with bands represent calculations from viscous hydrodynamic models with IP-Glasma initial state model~\cite{Schenke:2019pmk}. \label{fig:phenix_smallsys}}
\end{center}
\end{figure}

Recent results for high multiplicity events in small systems such as p+p and p+A collisions draw great attention because even in such small systems many similarities compared to the large systems have been reported, i.e. long-range correlation between two hadrons~\cite{Khachatryan:2010gv,CMS:2012qk,Abelev:2012ola,Aaij:2015qcq}, sizable flow coefficients $v_n$~\cite{Aad:2015gqa,Khachatryan:2016txc,PHENIX:2018lia,Sirunyan:2018toe}, and multiplicity scaling of (multi)strange hadron yields from large systems to the small system~\cite{ALICE:2017jyt}.
Figure~\ref{fig:CMS_ppRidge} shows two-particle correlation in high multiplicity p+p collisions at the LHC, where the long-range near-side correlation was observed along $\Delta\eta$ at $\Delta\phi\approx0$ which can be attributed to flow coefficients seen in large systems as discussed in Sec.~\ref{sec:hydro}.
Now a question is whether or not a QGP droplet is created. 
Viscous hydrodynamic models~\cite{Bozek:2011if,Weller:2017tsr} have been applied to explain the flow coefficients as shown in Fig.~\ref{fig:phenix_smallsys}.  The agreements between the model calculations and  the data are not as good as what we saw in Fig.~\ref{fig:v345-rhic},  which may imply differences in flow development in the A+A collisions and the small systems as mentioned in Sec.3.1.
A part of the difficulty and complexity in the interpretation comes from contributions from so-called non-flow, especially correlations of fragmented particles from back-to-back jets, which mimics the flow signal in small systems. 
Also, there is a discrepancy in the flow coefficients between PHENIX and STAR at this moment~\cite{PHENIX:2018lia,Lacey:2020ime}, which needs to be resolved before making conclusive remarks.

Strangeness enhancement in small systems is claimed to be explained assuming partial formation of QGP droplet~\cite{Kanakubo:2019ogh}.
On the other hand, jet quenching is not observed in the particle yield for small systems~\cite{Adam:2014qja,Aad:2016zif}, while a finite elliptic flow $v_2$ at high $p_T$ is observed in high multiplicity events of small systems~\cite{Aad:2019ajj} as in large systems, which could be explained by the path length dependence of parton energy loss. Possible contradiction between the two observables is still an open question and needs to be investigated.




%% file: 5-PhaseDiagram.tex
\subsection{Mapping of QCD phase diagram}

One of the main goals in heavy-ion collisions is to map out the QCD phase diagram (Fig.~\ref{fig:qcdPD}).
As mentioned in Sec.~\ref{sec:intro}, a first order phase transition at high $\mu_{\rm B}$ and a smooth cross over phase transition at small $\mu_{\rm B}$ are theoretically expected.
In Secs.~\ref{sec:eloss} and \ref{sec:hydro}, experimental results are shown that the QGP is formed in heavy-ion collisions at RHIC and the LHC corresponding to the region at large temperature with small $\mu_B$ in Fig.~\ref{fig:qcdPD} and the phase transition is considered to be the smooth cross over. 
In order to study the structure of the QCD phase diagram experimentally, one can vary the collision beam energy, which leads to the change in the temperature as well as baryon chemical potential $\mu_B$ in the reaction region~\cite{Bzdak:2019pkr,Fukushima:2020yzx}.


Assuming the fireball formed in the collisions is uniform and in chemical equilibrium, ratios of hadron yields can be well described by a simple statistical model with a few parameters; chemical equilibrium temperature $T_{\rm ch}$ and baryon chemical potential $\mu_{\rm B}$. 
Statistical hadronization models are found to remarkably describe the hadron yields for various particle species~\cite{Andronic:2017pug,Adamczyk:2017iwn,Becattini:2016xct}, indicating that the chemical equilibrium is achieved. The extracted $T_{\rm ch}$ and $\mu_{B}$ are plotted in Fig.~\ref{fig:qcdPD} and are close to the crossover critical temperature for hadronization predicted by lattice QCD at small $\mu_B$. As the collision energy decreases, $T_{\rm ch}$ decreases and $\mu_B$ becomes larger.

\begin{figure}[t]
\begin{center}
\includegraphics[width=0.8\linewidth,bb=0 0 221 373 ]{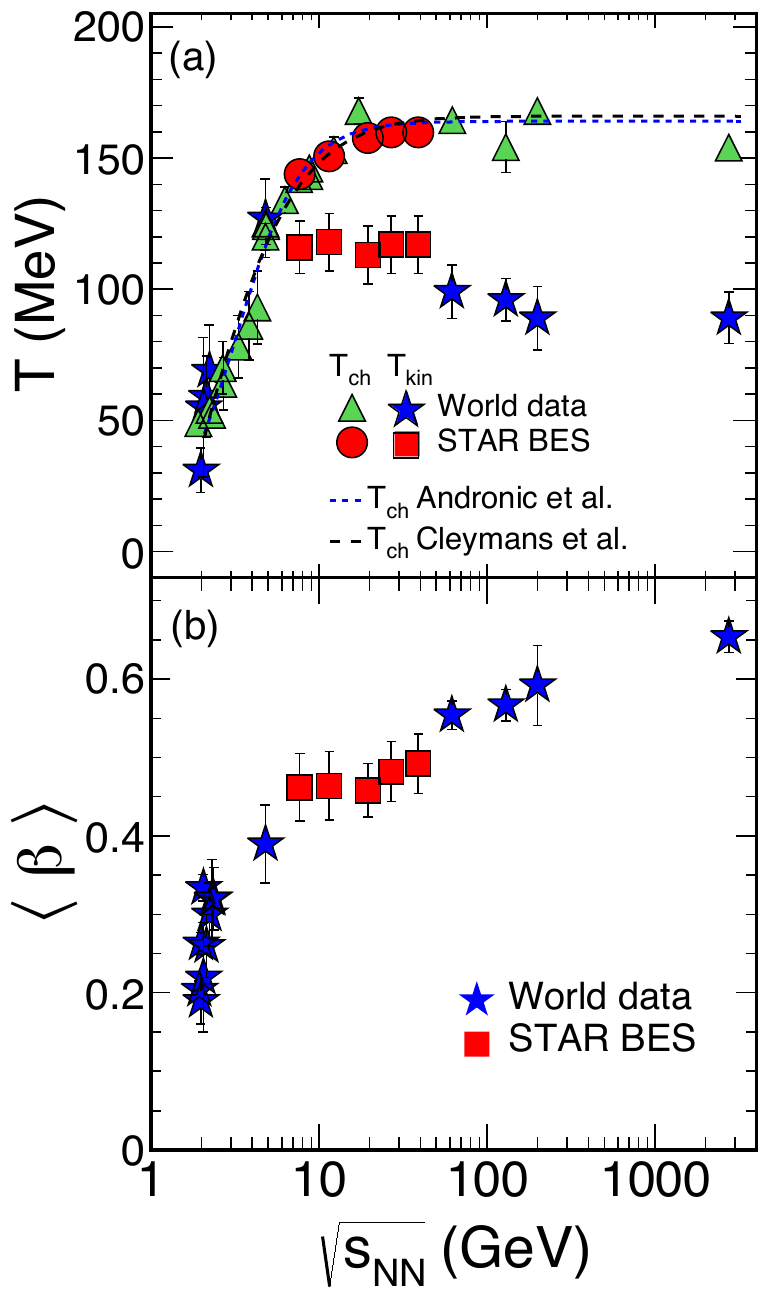}
\caption{Collision energy dependence of  (a) extracted chemical and kinetic freeze-out temperatures ($T_{\rm ch}$ and $T_{\rm kin}$) and (b) average transverse flow velocity ($\langle\beta\rangle$)~\cite{Adamczyk:2017iwn}. This figure is adapted from Ref.~\cite{Adamczyk:2017iwn}.\label{fig:TBeta}}
\end{center}
\end{figure}

As well as chemical equilibrium, kinetic freeze-out dynamics has been also studied employing a blast-wave model~\cite{Schnedermann:1993ws,Retiere:2003kf}. The model is based on the picture that particles are emitted from a boosted thermal source which is characterized by the freeze-out temperature $T_{\rm kin}$ and a common collective transverse flow velocity $\beta$. Systematic study of $T_{\rm kin}$ and average flow velocity $\langle\beta\rangle$ has been performed over a wide range of the collision energy as shown in Fig.~\ref{fig:TBeta}~\cite{Adamczyk:2017iwn}. The chemical freeze-out temperature $T_{\rm ch}$ stays constant above $\sqrt{s_{\rm NN}} \sim 10$~GeV, while it coincides with the kinetic freeze-out temperature $T_{\rm kin}$ around $\sqrt{s_{\rm NN}}=7$ GeV and sharply decreases when decreasing the energy. The flow velocity $\langle\beta\rangle$ becomes larger in the LHC energy, suggesting a strong expansion of the fireball, while it also shows a similar sharp drop around $\sqrt{s_{\rm NN}} < 7$ GeV.

\begin{figure}[htb]
\begin{center}
\includegraphics[width=0.8\linewidth,bb=0 0 466 737 ]{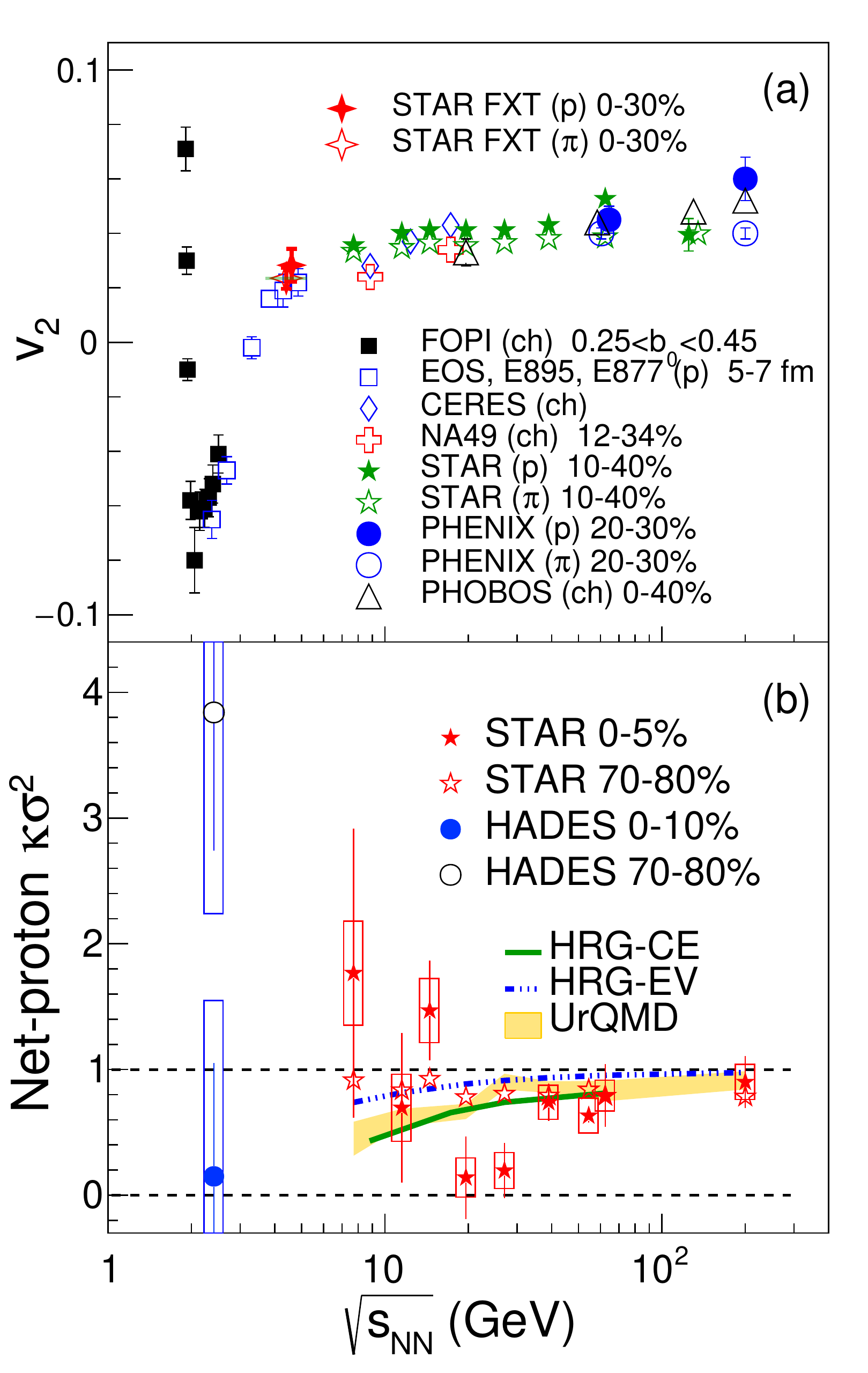}
\caption{Collision energy dependence of  (a) elliptic flow $v_2$~\cite{Adam:2020pla} (see Ref.~\cite{Adam:2020pla} and references therein for details) and (b) 4$^{\rm th}$-order cumulant ratio of net-proton distribution $\kappa\sigma^2$~\cite{Adam:2020unf,Adamczewski-Musch:2020slf}. \label{fig:V2C4}}
\end{center}
\end{figure}



The elliptic flow has been also extensively studied over a wide range of the collision energies as shown in Fig.~\ref{fig:V2C4}(a). The sign of $v_2$ changes twice in the plot; it changes from positive to negative and again becomes positive when the energy increases, corresponding to the rotational-like emission ($v_2>0$ at $\sqrt{s_{\rm NN}} \lesssim 1.4$ GeV), ``squeeze-out" emission due to spectator shadowing with an increased expansion of fireball ($v_2<0$), and the pressure-gradient-driven expansion ($v_2>0$ for $\sqrt{s_{\rm NN}}>4$ GeV) as discussed in Sec.~\ref{sec:hydro}.

These results also show a sudden change below $\sqrt{s_{\rm NN}}\approx8$ GeV, indicating change in underlying physics which might be related to the phase transition. Ongoing projects (STAR Beam Energy Scan Phase-I\hspace{-.1em}I including the fixed target program and HADES experiment) as well as future experiments (CBM at FAIR, MPD at NICA, CEE at HIAF, J-PARC-HI) are planned to study that energy range to search for a signature of the critical point.

\subsection{Critical point search}
Several observables have been proposed to probe a possible critical point in the QCD phase diagram, e.g. net-baryon fluctuations, directed flow, particle emission duration via femtoscopy, and neutron density fluctuation probed by light nuclei production (see Refs.~\cite{SN0696,SN0721} and references therein for details). 
Higher-order cumulants of net-proton distributions as a proxy for net-baryon number especially have been considered to be a promising tool to search for the critical point. 
The idea is based on the fact that the correlation length of fluctuations in net conserved quantities, such as net-charge or net-baryon number, diverges in the vicinity of the critical point.
Experimental results on net-proton higher-order cumulants shown in Fig.~\ref{fig:V2C4}(b) indicate non-monotonic behaviour over the collision energy~\cite{Adam:2020unf,Adamczewski-Musch:2020slf} but the uncertainty is still large. Also many other effects such as an experimental limitation of the measurement (acceptance and efficiency), finite size and lifetime of the system, baryon stopping, and non-equilibrium effect especially in lower energies, need to be understood before making a definitive claim. 

%% file: 6-Summary.tex
Extensive experimental and theoretical investigations have revealed that the matter produced in heavy-ion collisions at RHIC and the LHC is truly a new state of matter and the two features of the high opacity for color charges together with the non-viscous flow suggest that it is strongly-coupled QGP unlike expectations of almost non-interacting gas of deconfined quarks and gluons. 

On the other hand, we are still far from an understanding of the nature of the QGP phase transition: we do not know the location of the phase transition in terms of beam energies nor volumes of the fireball. We do not know whether the region of a first-order phase transition and the associated critical point is experimentally accessible. More experimental and theoretical studies are required.

Beam energies at AGS, SPS, RHIC to the LHC accelerators have been increasing steadily, but two directions are being considered for the future.  One direction is toward higher beam energies such as the CERN-FCC project~\cite{CERN-FCC} where higher temperature and longer lifetime are expected, where a more viscous fluid may be created and where we may have a chance to observe a weakly-coupled QGP.  
The other direction is to probe the vicinity of the phase transition point with a high luminosity, lower-energy accelerator.  From this viewpoint, new accelerator projects, NICA at JINR~\cite{Meshkov:2018mfl}, FAIR at GSI~\cite{FAIR}, CEE at HIAF~\cite{Yang:2013yeb}, and J-PARC-HI~\cite{JPARC-HI} have been proposed, some of which are under construction. These new facilities are the frontier of the field.

In addition, high energy collisions provide an unique opportunity to study hadron-hadron interactions, especially for unstable hadrons. Recent measurements on two baryon correlations, such as $\Lambda$-$\Lambda$~\cite{Lambda-Lambda_STAR} and $p$-$\Omega$~\cite{Acharya:2020klc}, in the small relative momentum of the pairs demonstrate that one can study the strong interaction between the baryons and search for exotic hadrons such as dibaryons. These measurements are a new tool providing valuable data for quantitative comparisons with Lattice QCD calculations and are also crucially important for the studies of neutron stars and hyperon interactions. The new frontiers, together with these new tools, will allow this field to continue
to grow over the next decades.
